\documentclass[12pt,preprint]{aastex}

\slugcomment{Submitted to The Astronomical Journal}

\shorttitle{New High Proper Motion Stars in the Northern Sky. I.}
\shortauthors{L\'epine et al.}

\begin{document}

\title{Spectroscopy of New High Proper Motion Stars in the Northern
Sky. I. New Nearby Stars, New High Velocity Stars, and an Enhanced
Classification Scheme for M Dwarfs.}

\author{S\'ebastien L\'epine\altaffilmark{1,2,3,4}, R. Michael
Rich\altaffilmark{2,5} and Michael M. Shara\altaffilmark{1}}

\altaffiltext{1}{Department of Astrophysics, Division of Physical
Sciences, American Museum of Natural History, Central Park West at
79th Street, New York, NY 10024, USA.}
\altaffiltext{2}{Visiting Astronomer, Lick Observatory.}
\altaffiltext{3}{Visiting Astronomer, MDM Observatory.}
\altaffiltext{4}{Visiting Astronomer, KPNO.}
\altaffiltext{5}{Department of Physics and Astronomy, University of
California at Los Angeles, Los Angeles, CA 90095, USA.}

\begin{abstract}
We define an enhanced spectral classification scheme for M dwarf
stars, and use it to derive spectral classification of 104 northern
stars with proper motions larger than $0.5\arcsec$ yr$^{-1}$ which
we discovered in a survey of high proper motion stars at low galactic
latitudes. The final tally is as follows: 54 M dwarfs, 25 sdK and sdM
subdwarfs, 14 esdK and esdM extreme subdwarfs, and 11 DA and DC white
dwarfs. Among the most interesting cases, we find one star to be the
coolest subdwarf ever reported (LSR2036+5059, with spectral type
sdM7.5), a new M9.0 dwarf only about 6pc distant (LSR1835+3259), and a
new M6.5 dwarf only 7pc from the Sun (LSR2124+4003). Spectroscopic
distances suggests that 27 of the M dwarfs, 3 of the white dwarfs, and
one of the subdwarfs (LSR2036+5059) are within 25pc of the Sun, making
them excellent candidates for inclusion in the solar neighborhood
census. Estimated sky-projected velocities suggest that most of our
subdwarfs and extreme subdwarfs have halo kinematics. We find that
several white dwarfs and non metal-poor M dwarfs also have kinematics
consistent with the halo, and we briefly discuss their possible
origin.
\end{abstract}


\section{Introduction}

The current census of stars in the Solar Neighborhood (the volume of
space within $\approx25$pc of the sun) is believed to be significantly
incomplete, especially at the faint end of the luminosity
function. Despite the fact that increasing numbers of substellar
objects (L dwarfs, T dwarfs), and low mass stars (M7-M9 dwarfs) are
now being discovered with the help of large infrared surveys
\citep{KRLCNBDMGS99, DTFEBFKT99, GR97} there remain significant numbers
of red dwarfs and white dwarfs which are still unaccounted for
\citep{HWBG97}.

Nearly all the local M dwarfs and white dwarfs are expected to be
brighter than magnitude $\sim20$ in the optical bands. The majority of
them should be detectable as stars with large proper
motions. If they haven't been identified yet it is because existing
all-sky surveys of high proper motion stars \citep{L79, L80} are
themselves significantly incomplete \citep{SIIJM00,
LSR02b}. Furthermore, the classification and characterization of even
the known high proper motion stars is still under way \citep{GR97,
JSML01, RKC02}.

Recently, several new additions to the solar neighborhood have been
confirmed from follow-up observations of newly discovered high proper
motion stars \citep{PGCDFBEFS01, HWBG02, SIILSS02a, RRSI02}. Follow-up
observations of faint stars with large proper motions have also
revealed the existence of previously unreported types of objects like
a very cool extreme subdwarf \citep{SSSIM99}, a magnetic DZ white dwarf
\citep{RLS01}, and a pair of nearby cool white dwarfs \citep{SSAII02b}.

In a previous paper \citep{LSR02b} we have reported the discovery of
140 new stars with proper motions larger than $0.5\arcsec$ yr$^{-1}$
at low galactic latitudes in the northern sky. For the past two years,
we have been carrying out a large spectroscopic follow-up survey of the
new high proper motion stars that we are finding in the northern
sky. The spectroscopy is being performed at the Lick Observatory, the
MDM Observatory, and the Kitt Peak National Observatory. This first
paper of a series presents spectral classification of a first set of
104 stars, all of which are listed in \citet{LSR02b}. Observations are
described in \S2. In \S3, we describe our classification method, which
expands on previous spectral index methods. Estimation of the radial
velocities and calculation of the spectroscopic distances is detailed
in \S4, where we also discuss the kinematics of the
objects. Especially interesting or intriguing stars are discussed in
\S 5. Important results are briefly summarized in \S 6.

\section{Spectroscopic observations}

Spectra were obtained on six separate runs, at the Lick Observatory in
August 2000, July 2001, and December 2001, at the MDM Observatory in
February 2001, and May 2002, and at the Kitt Peak National Observatory
in May 2002. All 64 Lick spectra were obtained with the KAST
spectrograph, mounted at the Cassegrain focus of the 3m Shane
Telescope. We imaged the stars through a 2.5$\arcsec$ wide slit, onto
the 600 l/mm grating blazed at 7500\AA\ . The KAST camera uses a
thinned CCD which displays significant fringing redwards of 7500\AA\ ,
and special care was needed to account for the fringing (using extra
dome flats), although some spectra appear to show residuals from the
fringing pattern on the far red side. The instrument rotator was used
to orient the slit vertically on the sky, to minimize slit loss due to
atmospheric diffraction. Standard reduction was performed with IRAF,
including sky subtraction, normalization, and removal of telluric
lines. The calibration was done using a set of KPNO spectrophotometric
standards \citep{MSBA88, MG90}. The resulting spectra cover the range
6200-9000\AA\ with a resolution of 2.34\AA\ per pixel.

The first 34 MDM spectra (February 2001) were obtained with the Mark
III Spectrograph mounted on the 2.4m Hiltner Telescope. Stars were
imaged through a 1.5$\arcsec$ slit, onto the 600 l/mm grating blazed at
5800\AA\ . We used a thinned CCD camera (called ``Echelle'') which
shows only slight fringing in the red. Standard reduction was
performed with IRAF, including sky subtraction, normalization, and
removal of telluric lines. For calibration, we used the same set of
KPNO spectrophotometric standards that we used for the Lick
observations (see above). The resulting spectra cover the range
5200-8800\AA\ with a resolution of 2.55\AA\ per pixel. Because of the
lower sensitivity of the setup to light redwards of 8000\AA\ , the
spectra are very noisy beyond that limit. Because of variable
conditions in some of the nights, removal of telluric lines was
sometimes difficult, and residual telluric absorption remains in some
of the spectra. On the second MDM run (May 2002) we used the MkIII
Spectrograph again, but this time with the 300 l/mm grating blazed at
8000\AA\ , which provided a clearer signal in the red. We also used a
thick CCD (``Wilbur'') to avoid problems with fringing. The 5 spectra
from the second run cover the range 6000-9900\AA\ with a resolution of
3.10\AA\ per pixel.

Finally, we observed the very faint white dwarf LSR2050+7740 during a
run on the Mayall 4m Telescope at the Kit Peak National
Observatory. We used the RC Spectrograph with the LB1A camera. Stars
were imaged through a 2.0$\arcsec$ slit onto the 316 l/mm grating blazed
at 7500\AA\ (\#BL181). The star was observed near the meridian with the
slit oriented north-south to minimize slit loss due to atmospheric
diffraction. Standard reduction was performed with IRAF, including sky
subtraction, normalization with the KPNO spectrophotometric standards,
and removal of telluric lines. The resulting spectrum covers the range
5800-9100\AA\ with a resolution of 1.74\AA\ per pixel.

\section{Spectral Classification}

\subsection{Spectral Indices and the Metallicity Scale}

A well-defined spectral sequence for M dwarfs in the red part of the
spectrum (6000\AA-9000\AA) was compiled by \citet{KHM91}. This sequence
has become the standard reference for the spectral classification of M
dwarfs. Classification on this system is typically performed by
fitting the spectrum of a star to this sequence, either by eye or
numerically through minimization methods. However, since the work of
\citet{RHG95}, spectral indices that measure the depth
of various molecular bandheads are being used to provide a
quantitative spectral classification of M dwarfs. Spectral indices
have been defined, and their values calibrated against spectral type,
by \citet{RHG95} for classification of M dwarfs in the Palomar-MSU
spectroscopic survey, by \citet{KHS95} for the classification of
late-type M dwarfs, by \citet{KRLCNBDMGS99} for the classification of
late-type M dwarfs and L dwarfs, by \citet{MDBGFZ99} also for the
classification of late-type M dwarfs and L dwarfs, and by \citet{H02}
for the classification of M dwarfs found in the Sloan Digitized Sky
Survey. A total of 45 different spectral indices are defined in the 5
papers quoted above. All of them are correlated with spectral type,
though some provide a better diagnosis than others.

A very useful classification scheme was developed by \citet[hereafter
G97]{G97}, based on four of the \citet{RHG95} spectral indices (CaH1,
CaH2, CaH3, TiO5). Several authors have performed spectral
classification using this system \citep{GR97, SSSIM99, JSML01,
CR02}. The G97 method separates red dwarfs into three metallicity
classes: dwarfs (dM or M), subdwarfs (sdM), and extreme subdwarfs
(esdM). It also yields a good spectral classification for sdM, esdM
and early/intermediate M dwarfs, but is not appropriate for the
classification of late-type M dwarfs because the four indices measure
the strength of molecular features around 7000\AA\ which tend to
saturate beyond spectral type M6. Classification of late-type M dwarfs
can be performed using the spectral indices defined in
\citet{KRLCNBDMGS99} and \citet{MDBGFZ99}, or by using the single
index defined by \citet{KHS95}. We maintain, however, that an improved
and unified scheme for the quantitative classification of all M dwarfs
is needed.

Here we expand on the G97 method, and define a set of
spectral indices which measure most major features in the
6000\AA-9000\AA\ spectral range (see Table 2). In Figure 1, we show
the spectral features measured by each individual index. From G97, we
adopt the CaH1, CaH2, CaH3 indices (which measure the strength of the
CaH bands blueward of 7050\AA) and the TiO5 index (which measures the
TiO bandhead redward of 7050\AA). From \citet{H02}, we
adopt the VO 7434 and TiO 8440 indices (renamed VO1 and TiO7,
respectively) which measure the strength of the VO band redward of
7530\AA\ and the TiO bandhead blueward of 8430\AA. We introduce two
new indices, TiO6 which measures the depth of the prominent TiO
bandhead redward of 7600\AA, and VO2, which measures the tail of that
same TiO bandhead blended with the VO band feature redward of 7850\AA\
which is very prominent in late-type M dwarfs. Our VO2 index is similar
to the VO 7912 index defined in \citet{H02}, but the
reference level (denominator) is defined around 8140\AA\ instead of
8410\AA ; we prefer to use a reference level that is closer on the
spectrum to the measured feature. Finally, we define a new Color-M
index which measures the slope of a pseudo-continuum using reference
points defined in narrow spectral ranges with a local maximum in
intensity. This color index is very useful in estimating the general
shape of the spectral energy distribution.


The values of the spectral indices were calculated for all the stars
using the SBANDS task in IRAF. Values obtained for all the M dwarfs,
subdwarfs, and extreme subdwarfs (i.e. excluding all the white
dwarfs), and listed in Table 3.


As a first step, we follow the prescription of G97 to separate our red
dwarf stars into the three metallicity classes. We assume that all the
stars in our list are gravitationally bound to the galaxy, which means
they cannot have a transverse velocity larger than 500 km/s. Because
our targets all have proper motions larger than 0.5 arcsec/yr, they
must be within 200 pc of the sun, which means a distance modulus
smaller than 6.5. Since all our stars have an apparent red magnitude
larger than 12, this means none of our stars can have an absolute
magnitude brighter than 5.5, and most are much fainter. This
completely rules out the possibility that any of these stars are
giants or supergiants. 

The separation into M, sdM, and esdM classes is based on the ratio of
the CaH band indices to the TiO5 index (see Figure 2). We use the
relationships defined in G97 for the separation of the stars
into the different classes; these relationships are illustrated on
Figure 2. To be considered an sdM/esdM, a star has to be within the
limits defined by G97 in at least two of the diagrams. The
low value of the CaH1 index for three late-type M dwarfs (TiO5 $<$
0.35) results from a low signal-to-noise ratio in those stars around
where CaH1 is measured. The CaH2/TiO 5 relationship was used to
separate the esdM from the sdM. For the late K dwarfs and early M
dwarfs, the G97 molecular bands criteria are more ambiguous;
the CaH indices and the TiO5 index all converge to 1 toward earlier
spectral types. There appear to be a number of stars with TiO5$>$0.7
that could be M dwarfs. However, these would have a spectral type
earlier than M2, which means they would have an absolute magnitude
M$_{R}<8$. However, if they are M dwarfs they would be too faint (R$>14$)
to be within a few hundred parsecs. Because of their large proper motions
($\mu>0.5\arcsec$ yr$^{-1}$) these M dwarfs would have extremely large
transverse velocities (V$_{T}>$500 km s$^{-1}$). For these reasons, it
is likely that these stars are relatively nearby subdwarfs, instead of
very distant, very high-velocity dwarfs.

\subsection{M dwarfs}

Our spectral types for the M dwarfs are assigned based on our expanded
classification scheme which uses all the indices defined in Table 2.
We have kept the relationships defined by G97 for the CaH2,
CaH3, and TiO5 indices. The spectral type $S_P$ correlates with
these indices as:
\begin{equation}
S_P = 7.91 \ {\rm CaH2}^2 - 20.63 \ {\rm CaH2} + 10.71 \ \ \ \ {\rm (M0-M6)}
\end{equation}
\begin{equation}
S_P = -18.00 \ {\rm CaH3} + 15.80 \ \ \ \ {\rm (M0-M6)}
\end{equation}
\begin{equation}
S_P = -9.64 \ {\rm TiO5} + 7.76 \ \ \ \ {\rm (M0-M6)}
\end{equation}
We give in parentheses the range over which these relationships are
valid. There are weaknesses to these indices which we now note, as a
prelude to an improved classification system given below. First, the
three indices above do not make use of the full spectral range
6000\AA-9000\AA\, where several strong molecular bands are found in M
dwarfs. Second, the CaH2, CaH3 and TiO5 indices are defined in a
low-intensity region of the optical spectral energy distribution (for
intermediate and late-type dwarfs). A third problem is that they are
not strictly independent, since CaH2, CaH3, and TiO5 all use
the same pseudo-continuum region as their reference point
(7042\AA-7046\AA). This reference region is only 4\AA\ wide, and is
therefore sensitive to spectral noise; any significant instrumental
deviation in the 7042\AA-7046\AA\ region propagates equally in all
three indices.

To derive the correlations for the other indices, we first obtained a
crude spectral typing by visual inspection of the spectra and
comparison with the \citet{KHM91} sequence. We then made fine
corrections to the spectral types (typically by no more than one
subtype), until we obtained a clean sequence where all indices were
linearly correlated with spectral type. The final sequence is shown in
Figure 3. The few outlying points in the graphs are associated with
stars for which we only have a noisy spectrum (especially
LSR0212+7012, LSR0200+5530, and LSR0646+3412). We used linear
regressions to obtain the following relationships:
\begin{equation}
S_P = -30.5 \ {\rm VO1} + 32.2 \ \ \ \ {\rm (M2-M8)}
\end{equation}
\begin{equation}
S_P = -11.2 \ {\rm TiO6} + 11.9 \ \ \ \ {\rm (M2-M8)}
\end{equation}
\begin{equation}
S_P = -10.5 \ {\rm VO2} + 12.4 \ \ \ \ {\rm (M3-M9)}
\end{equation}
\begin{equation}
S_P = -11.0 \ {\rm TiO7} + 13.7 \ \ \ \ {\rm (M3-M9)}
\end{equation}
Again we give in parentheses the range over which each of the
relationships is valid. The VO1, TiO6, VO2, and TiO7 relationships are
valid for later spectral types than are the CaH2, CaH3, and TiO5
relationships. The indices, taken together, can be used to obtain
spectral types over the whole M dwarf range to an accuracy of about
half a spectral type.

While the variation of all the spectral indices with spectral type
appears to be linear, there appears to be an exponential progression
of the Color-M index over the M dwarf sequence. In Figure 3, we plot
the logarithm of the Color-M index, and show that it can be fitted with
a linear relationship in the M3-M8 range. The Color-M uses two
spectral regions separated by more than 1500\AA\ and is thus not as
reliable as the other indices for spectral classification purposes, as
it is more sensitive to slit loss, extinction, and other calibration
errors. Nevertheless, the correlation is reasonably good, and the
Color-M index may be used to help determine the spectral type of
intermediate M dwarfs (M4-M8) using the relation:
\begin{equation}
Sp = 7.5 \ \log ({\rm Color-M}) + 1.6 \ \ \ \ {\rm (M4-M8)}
\end{equation}
The complete sequence of M dwarf spectra in displayed in
Figures 4-5. One can verify that there is a smooth, continuous
evolution toward later types.

We further examined the spectra for any indication of $H\alpha$
($\lambda6563$) in emission. Given the quality and resolution of the
spectra, we determined that $H\alpha$ detection could be considered
significant if it had an equivalent width smaller than about -2.0\AA
. We added the subscript $e$ to stars which qualified as $H\alpha$
emitters under this criterion. Close observation of the spectra
suggests that several more stars (including all those of spectral type
6.0 and later) may also be $H\alpha$ emitters, but the equivalent
width of the $H\alpha$ line was too weak to be considered
significant under the criterion above. High resolution observations of
the spectral region around 6563\AA\ would be required in order to
determine which stars can be considered to be emission-line stars in a
broader sense. It is remarkable that we find only 8 emission-line
stars in our sample of 54 M dwarfs, especially given the large number
of M5-M6 stars in the sample. Generally, about half of the M5-M6
dwarfs are found to be active \citep{GMRKLW00}. Because activity is
correlated with age, this possibly indicates that most stars in our
sample are relatively old M dwarfs. This would be consistent with the
relatively larger velocities that we find for most of them (see \S4.4
below).

\subsection{Subdwarfs and extreme subdwarfs}

For sdM stars, we simply follow the G97 classification system, which
uses only the CaH2 and CaH3 indices:
\begin{equation}
S_P = 7.91 \ {\rm CaH2}^2 - 20.63 \ {\rm CaH2} + 10.71 \ \ \ \ {\rm (sdK5-sdM7)}
\end{equation}
\begin{equation}
S_P = -16.02 \ {\rm CaH3} + 13.78 \ \ \ \ {\rm (sdK5-sdM7)}
\end{equation}
In this system, negative values of $S_P$ correspond to sdK stars, with
$S_P=-2\rightarrow$sdK5, and $S_P=-1\rightarrow$sdK7. We plot the
distribution of CaH1, CaH2, and CaH3 values as a function of the sdM
subtype in Figure 6. Because the TiO and VO bands are very weak in
subdwarfs, they cannot be used as classification criteria. The Color-M
term again appears to increase exponentially with the subtype, loosely
following the relationship:
\begin{equation}
Sp = 7.3 \ \log ({\rm Color-M}) - 0.6
\end{equation}
As expected, the slope of the pseudo-continuum measured by Color-M is
significantly shallower for a given subtype than it is for the M
dwarfs. The complete sequence of sdM spectra is displayed in Figure
7. One of our subdwarfs is classified with a very late spectral type
sdM7.5. To the best of our knowledge, this is the coolest subdwarf
ever identified.

For esdM stars, we also use the G97 calibration based on the
CaH2 and CaH3 indices.
\begin{equation}
S_P = 7.91 \ {\rm CaH2}^2 - 20.63 \ {\rm CaH2} + 10.71 \ \ \ \ {\rm (esdK5-sdM7)}
\end{equation}
\begin{equation}
S_P = -13.47 \ {\rm CaH3} + 11.50 \ \ \ \ {\rm (esdK5-sdM7)}
\end{equation}
Again, negative values of $S_P$ here correspond to esdK stars, with
$S_P=-2\rightarrow$esdK5, and $S_P=-1\rightarrow$esdK7. We plot the
distribution of CaH1, CaH2, and CaH3 values as a function of the sdM
subtype in Figure 8. The Color-M term is again correlated with
spectral type, following approximately the relationship:
\begin{equation}
Sp = 22.0 \ \log ({\rm Color-M}) - 4.1
\end{equation}
Many of our esdM spectra (especially those from the 2001 MDM run) are
relatively noisy in the red beyond 8000\AA\ . For this reason, the
Color-M index is not very reliable for those stars. The 3 points in
the bottom panel of Figure 8 that lie well outside the relation
defined in equation 14 are from noisy MDM spectra. We suspect that
the correlation between the esdM subtype and the Color-M index is
actually very well defined. Spectra obtained with minimal slit loss
and which are carefully normalized are, of course, required if the
Color-M index is to be used for the classification of esdM stars.
The complete sequence of esdM spectra is displayed in Figure 9.

\subsection{White Dwarfs}

The white dwarf stars in our sample are easily identified because they
show none of the molecular features observed in the red dwarf
stars. We compared our spectra to the catalog compiled by
\cite{WGLLFBG93}, and examined the spectra for the presence of
H$\alpha$ in absorption. The 11 white dwarfs in our sample are
either completely featureless (DC stars) or display some weak
H$\alpha$ emission (cool DA stars).

Following the definition that the numerical spectral subtype of white
dwarfs follows $50,400/T_{eff}$, we perform blackbody fits to our white
dwarf spectra in the 6000-9000\AA\ range. The corresponding blackbody
temperatures are given in Table 3. The fits are accurate to $\pm250$K,
and the spectral typing is accurate to one spectral subtype. All the
stars in our sample are late-type DA or DC white dwarfs, with DC7
being the earliest, and DC10 the latest. Our spectra for the white
dwarf stars are plotted in Figure 10. Because we used observational
setups optimized for the observation of red dwarf stars, our white
dwarf spectra are sometimes very noisy, and the classification is not
very reliable. The assignment of white dwarfs in the DC class is
particularly dubious, and may simply result from our inability to
detect the H$\alpha$ absorption line in noisy spectra. Still, our
spectra clearly show that all the white dwarfs discovered in the
sample are relatively cool, probably old objects.


\section{Distances and kinematics}

\subsection{Reduced proper motion diagrams}

A most useful tool for pre-selecting classes of high proper motion
stars is the reduced proper motion diagram. Recently, \citet{SG02}
have demonstrated that dwarfs, subdwarfs, and white dwarfs can be
separated out when the ratio of infra-red to optical colors is used in
making the reduced proper motion diagrams. The reduced proper motion
$H$ is obtained by adding the logarithm of the proper motion to the
apparent magnitude of the star \citep{L25}. Here we use two
different reduced proper motion terms: 
\begin{equation}
H_{\rm R} \equiv {\rm R} + 5 + 5 \log{\mu} = M_{\rm R} + 5 \log{v_t} -3.38
\end{equation}
\begin{equation}
H_{\rm K_s} \equiv {\rm K_s} + 5 + 5 \log{\mu} = M_{\rm K_s} + 5 \log{v_t} -3.38
\end{equation}
Because both the apparent magnitude (R, ${\rm K_s}$) and proper motion ($\mu$)
are a function of the distance to the star, the distance terms cancel,
and so the reduced proper motion really is a measure of the combined
absolute magnitude (M$_{\rm R}$, M$_{\rm K_s}$) and transverse velocity
($v_t$). Reduced proper motion diagrams are plots of the reduced
proper motion against a color term.

We have made reduced proper motion diagrams using [M$_{\rm R}$, B-R]
and [M$_{\rm K_s}$, R-${\rm K_s}$]. In Figure 11, we plot the stars for
which the corresponding magnitudes are available, using different
symbols for the different classes of objects: filled circles for M
dwarfs, open squares for sdM, filled triangles for esdM, and asterisks
for the white dwarfs. On the [M$_{\rm R}$, B-R] diagram, white dwarfs
populate a distinct locus, but it is not possible to distinguish the
dwarfs from the subdwarfs. On the [M$_{\rm K_s}$, R-${\rm K_s}$]
diagram however, the dwarfs, subdwarfs, and white dwarfs clearly
occupy different loci. Proper motion and optical+infrared photometry
alone can thus be used to guess what class of object one is dealing
with, although spectroscopic confirmation is still required for
certainty. This however, is a very useful method for pre-selecting
stellar populations for follow-up observations.

\subsection{Spectroscopic distance estimates}

To obtain a calibration of the absolute magnitude against spectral
type for M dwarfs, we used a set of nearby stars for which there exist
parallax measurements accurate to better than 10\%. We selected 55
stars from the NStars database (http://nstars.arc.nasa.gov) spanning a
range of spectral types from M0.5 V to M8.0 V. There were very few
objects in the M7.0 to M9.0 range, so we complemented the sample with
11 late-type dwarfs ranging from M6.5 V to M9.5 V whose parallaxes are
given in \citet{D02}. We proceeded to recover the B, R, and ${\rm
K_s}$ magnitudes of those stars {\em using the same sources as those
used for our sample of new high proper motion stars}, namely the Guide
Star Catalog-II (GSC2.2.1) for the B and R magnitudes, and the 2MASS
Second Incremental Release for the ${\rm K_s}$ magnitude. This
procedure guarantees that the stars can then be compared on exactly
the same magnitude system. While not all the stars had counterparts in
the GSC2.2.1 or 2MASS, we did obtain reference magnitudes for at least
35 stars in each band. We then used the parallax measurements to
derive a spectral-type / absolute magnitude calibration.

The correlations between spectral type and absolute magnitudes are
shown in Figure 12. the relationship is clearly not linear across the
M dwarf sequence, as the absolute B and R magnitudes drop quickly
between spectral types M4 and M6; the absolute K magnitude also
appears to drop faster between M3 and M5. We found that the overall
behavior of the correlations could be modeled using third order
polynomials. We performed ${\chi}^2$ minimization fits to obtain the
following empirical relationships between spectral types ($S_P$) and
absolute magnitudes:
\begin{equation}
S_P({\rm M}) = 0.0156 \ {\rm M_B}^3 - 0.730 \ {\rm M_B}^2 + 11.8 \ {\rm M_B} - 60.4
\end{equation}
\begin{equation}
S_P({\rm M}) = 0.0143 \ {\rm M_R}^3 - 0.531 \ {\rm M_R}^2 + 7.02 \ {\rm M_R} - 28.1
\end{equation}
\begin{equation}
S_P({\rm M}) = 0.0974 \ {\rm M_{K_s}}^3 - 2.076 \ {\rm M_{K_s}}^2 + 15.4 \ {\rm
M_{K_s}} - 35.9 .
\end{equation}
We then used equations 17-19 to derive reference absolute magnitudes
for each given spectral type $S_P$, and used these to obtain a
spectroscopic distance estimate for each of our M dwarfs. Results are
listed in Table 4. For stars with magnitudes in more than one band,
the average distance is used; distance estimates in the different
bands usually agreed to better than 30\%.

We use the same method to determine a spectral-type / absolute
magnitude calibration for sdM and esdM stars. We used as reference a
set of 25 sdM and 19 esdM stars with spectral type and astrometric
parallax given in G97. We looked for counterparts of these
stars in the Guide Star Catalog-II (GSC2.2.1), and the 2MASS
Second Incremental Release. The empirical spectral-type / absolute
magnitude relationships are plotted in Figures 13-14. In all cases,
the relationships appear to be linear. We thus performed
${\chi}^2$ minimization fit of first order polynomials to obtain the
following empirical relationships between spectral types ($S_P$) and
absolute magnitudes. For the subdwarfs (sdM), we find:
\begin{equation}
S_P({\rm sdM}) = 1.12 \ {\rm M_B} - 12.4
\end{equation}
\begin{equation}
S_P({\rm sdM}) = 1.13 \ {\rm M_R} - 10.0
\end{equation}
\begin{equation}
S_P({\rm sdM}) = 1.65 \ {\rm M_{K_s}} - 10.4 .
\end{equation}
And for the extreme subdwarfs (esdM), we find:
\begin{equation}
S_P({\rm esdM}) = 1.85 \ {\rm M_B} - 23.2
\end{equation}
\begin{equation}
S_P({\rm esdM}) = 2.05 \ {\rm M_R} - 21.4
\end{equation}
\begin{equation}
S_P({\rm esdM}) = 2.68 \ {\rm M_{K_s}} - 21.6 .
\end{equation}
Distances for the sdM and esdM stars in our sample are based on these
empirical relationships. Estimated values are given in Table 4.


Errors on the distance estimates depend on three main factors. First
of all, there is an uncertainty related to the scatter in the
spectral-type/absolute magnitude relationships. This scatter arises
because of errors in the magnitudes and/or astrometric parallax, but
may also result from an intrinsic scatter in the empirical
relationship. There is typically a 0.5 magnitude scatter in the
spectral-type/absolute magnitude relationship which translates into a
distance error of about 25\%. Furthermore, there is also an error of
about 0.5 mags in the POSS-II photographic $B$ and $R$ magnitudes,
which should also result in an extra 25\% error on the distance
estimate. The error is much smaller for the 2MASS $K_s$ magnitude, but
we do not have $K_s$ values for all the stars. The fact that we use a
combination of the spectral-type/absolute magnitude relationships for
the $B$, $R$ and $K_s$ somehow reduces the errors on the distance
estimate. Nevertheless, there should be an basic uncertainty on all
the distances of M dwarfs and subdwarfs of about 35\%.

Second, there is a significant difference between the calibrations of
the different metallicity classes. This is especially true for early
type stars, where for example an M0V star is fully 3 magnitudes
brighter in the B and R bands than an esdM0 (see Figs. 12-14). While
it is easy to distinguish dwarfs and extreme subdwarfs, the separation
between dwarfs and subdwarfs, and between subdwarfs and extreme
subdwarfs is arbitrary, and one actually expects to find a continuous
distribution of stars over the whole metallicity range. The rough
separation into only three metallicity classes thus introduces errors
in the distance estimates. A star lying at the limit between two
spectral classes can fall up to one magnitude from one or another
spectral-type/absolute magnitude relationship, which means its
distance estimate could be up to 60\% in error.

A third factor is the possibility that some of the stars are actually
unresolved binaries. Distances of unresolved binaries tend to be
underestimated as the pair is intrinsically brighter for its apparent
spectral type. The effect is largest for two M dwarfs of equal
luminosity (and mass), where the apparent distance is underestimated by
a factor $\sqrt{2}$, or about 40\%.

Distance estimates for M dwarfs should be relatively more accurate
than for the subdwarfs and extreme subdwarfs. The M dwarfs are better
defined as a class, and their spectral-type/absolute magnitude
relationship is supported by a larger number of astrometric standards
(67 stars). Our relationships for sdM and esdM stars, on the other
hand, are derived using fewer astrometric standards (25 stars for the
sdM, and only 19 for the esdM). All things considered, our distance
estimates {\em for individual objects} are probably accurate to about
50\% for the M dwarfs, and to a factor of two for the sdM and esdM.


For the white dwarfs, we obtain very crude photometric distances with
the B-R color, by using the relation given in \citet{OHDHS01}:
\begin{equation}
{\rm M_B} = 12.73 + 2.58 ({\rm B - R}) .
\end{equation}
For those stars for which we do not have B magnitudes, we compare them
with other white dwarfs in our sample with similar spectral types and
estimated their relative distances based on their relative R
magnitudes. Photometric distances for the white dwarfs are listed in
Table 1. The accuracy of the distance estimates is very dependent on
the accuracy of the $B-R$ color term. Because we use photographic $B$
and $R$ magnitudes, large errors (possibly up to a factor of 2) can be
expected in the distance measurements.


\subsection{Radial velocities}

Our medium resolution (2-3\AA\ per pixel) spectra were initially
obtained for the purpose of classification alone, and are not ideally
suited for measuring radial velocities. Nevertheless, we find that
radial velocities can still be obtained with typical errors of
$\pm30-50$km s$^{-1}$. While relatively large, these errors are
nevertheless reasonably small when one deals with old disk or halo
stars, which can have radial velocities of a few hundred km
s$^{-1}$. Because our sample of high proper motion stars is strongly
biased towards stars with very large velocity components, we naturally
expect many of the stars to also have very large radial velocities.

We estimate radial velocities for each of our red dwarfs and subdwarfs
by measuring the centroids of strong atomic absorption lines. In
subdwarfs and extreme subdwarfs, we use two different sets of lines
depending on the available wavelength coverage. For the 2001 MDM
spectra, more sensitive at shorter wavelengths, we use the \ion{Na}{1}
$\lambda$5889.9 and $\lambda$5895.9 doublet and the \ion{Ca}{1}
$\lambda$6122.2 and $\lambda$6162.7 lines. For the Lick spectra and
the 2002 MDM spectra, more sensitive at longer wavelengths, we use the
\ion{K}{1} $\lambda$7664.9 and $\lambda$7699.0 lines and the
\ion{Ca}{2} $\lambda$8542.1 and $\lambda$8662.1 lines. For the M
dwarfs, all radial velocity measurements are based mainly on the
centroids of the \ion{K}{1} $\lambda$7664.9 and $\lambda$7699.0
lines. In emission line stars, we also measured the centroid of the
$H\alpha$ emission line. It was not possible to obtain radial
velocities from the featureless DC white dwarfs and for 2 of the DAs
with very weak and noisy $H\alpha$ lines. However, we have obtained
radial velocity estimates for 3 DA white dwarfs, by measuring the
centroid of the core of the $H\alpha$ absorption.

For all the stars in the MDM runs, we measured the wavelength shifts
relative to the XeNeAr comparison arcs obtained after each
exposure. For the Lick spectra, sky lines from each individual
exposure were used as a reference. Heliocentric radial velocities
(HRV) were calculated after applying the relevant corrections. The
resulting values are listed in Table 1.


\subsection{Kinematics}


We derive UVW components of the space velocity for each individual
object using the proper motion, spectroscopic distance estimate, and
radial velocity. We take the U component to be the velocity
toward the galactic center, V the velocity toward the direction of
galactic rotation, and W the velocity toward the north galactic
pole, all calculated relative to the local standard of rest
(l.s.r.). We use for the motion of the Sun relative to the l.s.r. the
empirically estimated values $[U,V,W]=[+10,+5,+7]$km s$^{-1}$
\citep{DB98}. We plot the distribution in the VU, VW and UW planes in
Figure 15 for the M dwarfs, in Figure 16 for the white dwarfs, and in
Figure 17 for the subdwarfs and extreme subdwarfs.

The velocity distribution of the different classes of objects are
compared to the observed distribution of local weakly metal-poor and
strongly metal-poor stars \citep{CB00}. The dotted lines in Figs.15-17
mark the observed 2$\sigma$ dispersion of a large sample of solar
neighborhood stars with [Fe/H]$\simeq$-0.5, which are associated with
the galactic disk. Based on \citet{CB00}, we hence adopt a velocity space
distribution for disk stars centered on [$U,V,W]=[0,-30,0]$km
s$^{-1}$, with a dispersion
$[2\sigma_U,2\sigma_V,2\sigma_W]=[100,110,60]$km s$^{-1}$. The dashed
lines in Figs.15-17 mark the observed 2$\sigma$ distribution of
[Fe/H]$\simeq$-2.5 stars, and associated with the galactic halo. For
halo stars, we adopt a velocity space distribution centered on
[$U,V,W]=[0,-190,0]$km s$^{-1}$, with a dispersion
$[2\sigma_U,2\sigma_V,2\sigma_W]=[280,220,190]$km s$^{-1}$, again
based on the distribution of very metal-poor stars reported by
\citet{CB00}.

The observed distribution in UVW velocity space for our M dwarfs
(Fig.15) shows a relatively large scatter, significantly larger than
expected for non-metal poor disk stars. Assuming that non-metal poor M
dwarfs should have the same kinematics as more massive stars, over
95\% of the local M dwarfs are expected to fall within the 2$\sigma$
range illustrated in Fig.15. We find, however that about 40\% of our
stars fall outside that range. Two major effects possibly
compound to yield the large velocity dispersion of our sample. First,
the errors on the individual UVW components are relatively large,
since they depend on the spectroscopic distance measurements and
radial velocities, both of which have relatively large
uncertainties. It is possible that more accurate measurements would
yield a smaller scatter in the velocity distribution. Second, and most
important, these stars have been selected based on their having large
proper motions, which means that our sample is strongly biased in
favor of stars with unusually large velocities.

Even if we account for the relatively large errors in the UVW
estimates, our sample clearly contains a significant number of M
dwarfs whose kinematics is more consistent with a halo membership (see
Fig.15). Hence we conclude that without a doubt {\em there exist non
metal-poor M dwarf stars with halo-like kinematics around the
neighborhood of the Sun}. Since it is unlikely that these stars were
actually born in the halo, there must be a mechanism that is
responsible for sending disk stars into halo-like orbits. Dynamical
ejection during promiscuous encounters of binary and single stars in
star clusters is one possibility \citep{HS02}, although these are
arguably relatively rare events. These low-mass, non metal-poor stars
with unusually large velocities could also have originated in very
massive star clusters, where the rapid expulsion of cluster gas by
massive stars can result in the evaporation from the cluster of less
massive members \citep{K02}. Alternatively, these stars could have
been accreted from a satellite galaxy in a merger event with the Milky
Way. Another possibility is that these stars are actually bulge stars
that have been sent into very elliptical orbits through gravitational
perturbations in dense regions of the bulge \citep{RGMFU98}. Our
current sample of M dwarfs with halo-like kinematics is however
relatively small at this point. Clearly, follow-up observations of
more faint high proper motion stars are required to discover more of
the non-metal poor, high velocity stars, and better characterize their
group kinematics.

Two of the stars in our sample of M dwarfs have extremely large
velocity components which place them well outside the 2$\sigma$ limit
of very metal-poor (halo) stars. These are LSR1844+0947 (M2.0V) and
LSR2010+3938 (M1.5V). These happen to be the two earliest stars in our
sample of M dwarfs. To find non-metal poor stars with such
unbelievably large velocities raises suspicions about their estimated
distances and radial velocities. In particular, those stars could be 
misclassified subdwarfs, in which case they could be much closer and
their velocities would then fall within the normal range of metal-poor
stars. We indicate this possibility by labeling LSR1844+0947 as
(sd)M2.0V, and LSR2010+3938 as (sd)M1.5V. In Table 4, we give in
parenthesis the estimated distances and velocity components for both
stars under the assumption that they are subdwarfs.

The velocity space distribution of our 11 white dwarfs is plotted in
Figure 16. Though we have very few objects, and the errors on the
distances (and thus on the velocity components) are relatively large,
it does appear that the white dwarfs have a velocity space
distribution that is very similar to that our our M dwarfs, with most
of them falling within or just outside the limits of disk stars, and a
few outliers with halo-like kinematics. While it is tempting to argue
that the white dwarfs with the largest space velocities are actually
members of the population of old halo white dwarfs claimed by
\citet{OHDHS01}, it is not possible at this point to exclude the
possibility that the high velocity white dwarfs are drawn from the
same population as those high velocity, non metal-poor M dwarfs. If
so, these white dwarfs would not have been born with halo kinematics,
rather acquiring them through some dynamical mechanism, conceivably
the one that put large numbers of non metal-poor M dwarfs on halo-like
orbits (see above).

The velocity space distribution of our subdwarfs and extreme
subdwarfs (Fig.17), on the other hand, is largely consistent with a
population of true halo stars. A significant number of stars are found
outside the 2$\sigma$ limits of the very metal poor stars. This again
may be due to the large errors in our velocity estimates (especially
considering the possible factor 2 error on the spectroscopic
distances) or to proper motion selection effects.

The velocity of one subdwarf is so large as to suggest that this star
is not gravitationally bound to the galaxy. The star LSR0400+5417,
classified as an sdK7 has an estimated velocity relative to the local
standard of rest $>1300$km s$^{-1}$ (it cannot be seen in Fig 17, as
it falls outside the limits of the plots). This star could be an
actual extra-galactic object, but it is more likely that the large
apparent velocity can be explained by an overestimated distance. We
suggest that this star may actually be a misclassified extreme
subdwarf, which we indicate by assigning the label (e)sdK7. In Table
4, we show in parenthesis the estimated distances and velocity
components assuming the star is an esdK. The distance is reduced by
more than a factor of two, and the star falls back into a
gravitationally bound galactic orbit. This illustrates well how errors
in the spectral classification can dramatically affect the estimated
distance and kinematics of M subdwarfs. A few other stars fall just
beyond the limit of gravitationally bound orbits, but again this most
likely reflects overestimated distances. Indeed it is possible that
all the stars that lie beyond the 2$\sigma$ limit of very metal-poor
(halo) stars may have overestimated distances. As it turns out, most
of these are currently classified as sdK/sdM (see Fig.17); if it turns
out that these are also misclassified esdK/esdM stars, then they may
in reality have velocities within the 2$\sigma$ limit of halo stars.

At this point, we see no apparent difference between the velocity
space distribution of the sdM and esdM stars, but any specific
difference in their velocity distribution might be blurred by the
large errors in the UVW components. In any case our sample is
relatively small, and a more extensive study of the velocity
distribution of a larger sample of low-mass subdwarfs and extreme
subdwarfs would be most instructive.


\section{Notes on Individual Objects of Interest}

\subsection{LSR0200+5530: a non-metal poor M dwarf with very large
space motion}

With a magnitude R=19.7, LSR0200+5530 is the faintest star in our
sample. While our MDM spectrum for that star is particularly noisy, it
strongly suggests a spectral type M5.5 V; the TiO bands are clearly
too strong for the star to be a subdwarf. This then places the star at
a distance d=90pc, with a surprisingly large space motion for a non
metal-poor star (U=-120 km s$^{-1}$, V=-200 km s$^{-1}$). Thus,
LSR0200+5530 appears to be a solar metallicity star, or perhaps a
marginally metal poor star, on a halo-like orbit. Since it is unlikely
that a non metal-poor star would originate in the halo itself, this
star appears to be a clear example of a star born in the disk that
later migrated to the halo. A close gravitational interaction with a
binary in a dense star cluster, or evaporation from a massive cluster
may have sent this one flying away.

\subsection{LSR0400+5417: an extra-galactic star?}

It is unclear whether LSR0400+5417 is a subdwarf (sdK) or an extreme
subdwarf (esdK) star. While its spectrum is consistent with an sdK7
subtype, the inferred spectroscopic distance is very large (d=375pc)
and suggests an enormous velocity relative to the
l.s.r. ($\approx$1300 km s$^{-1}$). If this is real, then this metal
poor star is an extragalactic object that just happens to be moving
through the Galaxy at this time. On the other hand LSR0400+5417 may
actually be an intrinsically fainter esdK7 star at a much closer
distance. Based on the extreme subdwarf distance calibration, we
estimate that as an esdK7, this star would be at d=150pc. This
reduces the space motion to $\approx$550km s$^{-1}$, which is still
extremely large, but is at least consistent with the star being on an
very eccentric, bound halo orbit.

\subsection{LSR0419+4233: an ultra-cool M8.5V dwarf on a halo-like
orbit}

With a spectral type M8.5 Ve, this low-mass star is at a spectroscopic
distance of only about 10pc. This star, however, has a very large
proper motion ($\mu=1.535\arcsec$ yr$^{-1}$) which at a distance of
10pc translates into a velocity $>90$km s$^{-1}$ relative to the local
standard of rest). This space motion is unusually large for a
non-metal poor disk star, and one has to wonder if the star might not
actually be much closer, which would reduce its estimated velocity. At
a distance of 10pc or larger, its velocity suggest that it might be a
low-mass, local representative of the thick disk. This is clearly a
high priority target for follow-up parallax measurements.

\subsection{LSR0539+4038: a new nearby M8.0Ve dwarf.}

This ultra-cool M8.0 Ve dwarf is estimated to be at a distance of only
10pc. Its large proper motion ($\mu=1.057\arcsec$ yr$^{-1}$) is
clearly a result of its proximity. Again, this is an excellent
candidate for inclusion in the census of very nearby stars.

\subsection{LSR0549+2329: a white dwarf with halo-like kinematics?}
 
This cool DC8 white dwarf appears to have a very large space motion
typical of a star with halo kinematics: at least 260 km/sec and
possibly larger, as the radial velocity is not measurable. This holds
true even if we assume that its distance has been overestimated by up
to a factor of 2. While this star might be a member of the
hypothetical population of very old halo white dwarfs suggested by
\citet{OHDHS01}, the fact that we are finding non-metal poor M dwarfs
with halo kinematics (e.g. LSR0200+5530 above) suggest instead that
this white dwarf was possibly born in the disk and eventually migrated
to the halo.

\subsection{LSR0556+1144: a bright, previously unreported, nearby dwarf}

This relatively bright R=14.2 star is the second brightest of our
sample. With a spectral type M5.5 V, it is also relatively cool and
intrinsically faint, and has a spectroscopic distance of only
10pc. This is a good candidate for inclusion in the census of very
nearby stars. The fact that this star is in a relatively crowded,
low-galactic latitude field explains why it could have been missed in
previous nearby star searches.

\subsection{LSR1722+1004: a non-metal poor M dwarf ejected from the disk?}

This is another M dwarf with disk-like abundances and halo-like
kinematics. Its spectrum is clearly that of a metal-rich M4.0 V, but
its spectroscopic distance of d=70pc implies a very large space motion
($\approx$250km s$^{-1}$). The UVW velocity components suggest that
this star is on a highly eccentric galactic orbit that brings it into
the halo. Again, this could be a star initially born in the disk that
later got ejected out to the halo.

\subsection{LSR1802+0028: an extreme subdwarf on a very eccentric,
retrograde halo orbit}

This extreme subdwarf is very distant (spectroscopic distance
d=180pc), and it has an extremely large velocity relative to the local
standard of rest, with an estimated V=-640 km s$^{-1}$. Even if its
distance has been overestimated by a factor two (which would imply
that this star fall on a sequence below that of even the extreme
subdwarfs), it would still have V$<$-500 km s$^{-1}$ because of its
extremely large radial velocity. There is little doubt that this
exceptional object is a true halo star. Its spectrum actually shows
very little trace of molecular TiO absorption. This most likely is an
extremely metal poor star on a very eccentric, retrograde halo orbit.

\subsection{LSR1833+2219: a companion to GJ718}

This star is a companion to the nearby star GJ718A, also known as the
flare star V774 Her. In a previous paper \citep{LSR02a}, we named this
star GJ718B, and reported a spectral type M4.5 V. This spectral type
was based on the G97 indices alone. The use of our expanded system
yields a spectral type M5.0 V instead. The spectroscopic distance of
26pc is very close to the actual distance of 23.4pc measured by
HIPPARCOS for GJ718A \citep{E97}. This supports the idea that our
spectroscopic distances are reasonably accurate, at least for the M
dwarfs.

\subsection{LSR1835+3259: an M9.0V star only 6 parsecs from the Sun}

With a spectral type M9.0 V, LSR1835+3259 is the coolest star in our
sample. Its optical and infrared magnitudes all are consistent with
the star being at about 6pc from us. We note that only two other M9.0
dwarfs are known to be within 7pc of the sun: DENIS 1048-39 at a
distance of about 4pc, and LP944-20 at a distance of about 5pc
\citep{DFMGBCAEFST01}. This star is a prime candidate for follow-up
astrometric parallax measurement.

\subsection{LSR1914+2825AB: a pair of high velocity subdwarfs}

This pair of early type subdwarf (sdM) stars is very peculiar because
of their large spectroscopic distance. Assuming a distance of 200pc, the
pair has a huge space motion of $\approx$500km s$^{-1}$ relative to
the local standard of rest, placing it on a halo orbit. The
angular separation between the components of $\approx12\arcsec$
implies a current orbital separation $>2400$AU.

The 40\% difference in the estimated distance of the two stars
illustrates the relatively large errors that plague the distance
estimates of sdM stars. Because the two stars most probably have the
same metallicity, one might expect that errors related to the
metallicity class assignment would be systematic, and that the
relative estimated distances of the two stars should be relatively
small. However, because the slope of the spectral-type/absolute
magnitude calibration changes with metallicity (see Figs.13-14),
errors on the relative distances may still occur because of the crude
assignment in a monolithic metallicity class.

On the other hand, it is also possible that LSR1914+2825B is an
unresolved binary. In this case, its distance might be underestimated
by up to 40\%, which would nicely explain the observed mismatch in the
relative distances.

\subsection{LSR1928-0200AB: a pair of M dwarfs in the halo?}

This is a pair of M dwarfs with a large orbital separation. The
earlier M3.5 V component LSR1928-0200A has a spectroscopic distance of
85pc, and the later M5.5 V component LSR1928-0200B has a spectroscopic
distance of 65pc. Their 10\arcsec angular separation translates into a
current orbital separation $\gtrsim650$ AU. While the 30\% relative
distance is not inconsistent with our estimated errors, it is possible
that LSR1928-0200B might be an unresolved binary, with its distance
somewhat underestimated. In any case, the relatively large distance of
the pair translates into an extremely large space motion
($\approx$280-360 km s$^{-1}$ to the local standard of rest) that is
atypical of non metal-poor stars.

The existence of such a system is quite puzzling. How can a non
metal-poor, or slightly metal poor binary have halo-like kinematics? A
very complex multi-body dynamical event in a cluster cannot be ruled
out, but would be unlikely. A more reasonable explanation might be the
evaporation of the system from a very massive cluster, in the scenario
described in \citet{K02}. Another possibility is that this system has
been accreted by the Galaxy, possibly through a merger event with a
slightly metal poor galaxy.

\subsection{LSR1945+4650AB: a pair of cool white dwarfs}

This is a pair of cool white dwarfs (DA9+DC10) for which we are having
some problem in estimating their distance. The photometric distance
derived for LSR1945+4650A (30pc) is half the photometric distance of
its companion (60pc). The two stars are a clear common proper motion
pair, and the chances are extremely small that they are not physically
related. With a $\approx9\arcsec$ angular separation, the two stars
are probably several hundred AU apart. The apparent discrepancy in the
estimated photometric distances rather suggests that the method is
prone to relatively large errors. The photometric distance is based on
the B-R color term (Equation 26). Both B and R are derived from
photographic plates, and accurate to only about 0.5
magnitudes. Therefore, B-R is accurate to about 0.7 magnitudes, which
can lead to a 40\% error in the distance estimate of a single object,
which means that there typically is a factor 2 error in the relative
estimated distance between two stars. It is also possible that
LSR1945+4650A is itself a pair of spatially unresolved white dwarfs,
in which case our distance of 30pc could be significantly
underestimated.

\subsection{LSR2036+5059: the coolest known subdwarf}

We find for this star a spectral type sdM7.5, which is unusually late
for a subdwarf. As far as we know, LSR2036+5059 is the coolest M
subdwarf reported to date. The coolest subdwarf reported in the
literature so far was LHS377, with a spectral type sdM7.0
\citep{G97}. The spectrum of LSR2036+5059 is distinctly odd-looking with
its very deep CaH band and relatively weak TiO features (see Figure
7). With a transverse velocity of $\sim$90km s$^{-1}$, it is a
possible member of the old-disk population. This star could have a
mass extremely close to the hydrogen burning limit for metal poor
stars. At a spectroscopic distance of only 18pc, this subdwarf is also
a very likely member of the Solar Neighborhood.

\subsection{LSR2122+3656: a very cool, extreme subdwarf}

This star, with spectral type esdM5.0, is the coolest extreme subdwarf
in our sample. The coolest extreme subdwarf known to date is the esdM7
star APMPM J0559-2903 \citep{SSSIM99}. LSR2122+3656 is also the closest esdM
in our sample, at a spectroscopic distance of 45pc. This emphasizes
the extreme rarity of the esdM stars in the Solar Neighborhood, and
shows the effectiveness of high proper motion surveys in recovering
them.

\subsection{LSR2124+4003: an M6.5V star only 7 parsecs from the Sun}

This cool red dwarf (M6.5 V) was among the brightest stars in our
sample (R=14.9). With a spectroscopic distance of about 7pc, it may
prove to be among the 100 nearest systems. Along with LSR1835+3259,
this is a top candidate for follow-up, astrometric parallax
measurements.

\section{Conclusions}

We have obtained spectra for 104 new stars with large proper motions
($\mu>0.5\arcsec$ yr$^{-1}$) found at low galactic latitudes
\citep{LSR02b}. We find that 54 of the targets are M dwarfs (M), 25
are subdwarfs (sdK, sdM), 14 are extreme subdwarfs (esdK, esdM), and
11 are late-type DA and DC white dwarfs.

We have expanded and refined the G97 classification method for M dwarfs
and subdwarfs by defining a larger set of spectral indices whose
calibration with spectral type can be used for spectral
classification. The new scheme can be applied to perform spectral
classification of all M dwarfs (early and late-type), subdwarfs, and
extreme subdwarfs, from a spectrum covering the 6000\AA-9000\AA\
wavelength range. Among the M dwarfs classified in this paper, 8 have
a spectral subtype M7 and later, including one new M9.0V star. We also
find one subdwarf with a very late spectral type of sdM7.5, the
coolest subdwarf ever reported.

We have provided a crude classification of the 11 white dwarf stars,
from blackbody fits of the spectral energy distribution in the
6000\AA-9000\AA\ range. All the white dwarfs are relatively cool DA
and DC, with the warmest at DC8 (6500K) and the coolest at DC11
(4750K). Many of the DC spectra are relatively noisy, and the DC
spectral class was assigned only by default, for lack of clear
identification of H$\alpha$ absorption. We thus suspect some of our
DCs might be actual cool DAs.

We have derived a spectral-type / absolute magnitude calibration
using sets of M, sdM, and esdM stars with published astrometric
parallaxes. We have found the counterparts of those stars in the
Guide Star Catalog-II (GSC2.2.1) and in the 2MASS Second
Incremental Release, to obtain their observed B, R, and ${\rm K_s}$
magnitudes. The spectral-type / absolute magnitude relationships for M
dwarfs was modeled with a third order polynomial, while we fit a linear
relationship for the sdM and esdM stars. Using the empirically
determined spectral-type / absolute magnitude relationships, we have
determined spectroscopic distances for all the observed red dwarfs and
subdwarfs in our sample.

We have obtained crude radial velocity measurements for most of our
stars using centroid measurements of \ion{Na}{1}, \ion{K}{1},
\ion{Ca}{1}, and \ion{Ca}{2} atomic absorption lines. Combining these
radial velocities with the estimated distances and proper motions, we
estimate the UVW velocity components (measured relative to the local
standard of rest) of all the stars. The distribution of M dwarf stars
in UVW space is consistent with most of the stars being components of
the disk. However, some have velocities that are more consistent with
halo orbits, which is quite surprising for non-metal poor stars. We
suggest that these stars originate in the disk but got ejected into
halo orbits through some yet undefined mechanism. Possible scenarios
include 3-body gravitational interactions or massive cluster
evaporation.

While we find the UVW space distribution of white dwarfs to be similar
to that of the M dwarfs, the velocity space distribution of subdwarfs and
extreme subdwarfs is markedly different, and is largely consistent
with the distribution of known, very metal-poor, candidate halo stars.
We find that a significant number of sdM and esdM stars fall outside
the 2$\sigma$ limit of halo stars in UVW space. These stars are
apparently on very eccentric halo orbits, some of which appear to
be at the limit of being gravitationally bound to the Galaxy.

Finally, we report the discovery of a number of new, very nearby
stars, including two stars that are most likely to be well within 10pc
of the Sun, and 29 more (25 M dwarfs, 3 white dwarfs, and 1 subdwarf)
that are likely to be within 25pc of the Sun. The two new very nearby
stars are LSR1835+3259, an M9.0 dwarf at about 6pc, and LSR2124+4003,
and M6.5 dwarf at about 7pc from the Sun.

Follow-up spectroscopy of newly discovered faint stars with large
proper motions proves yet again to be very productive in recovering
intrinsically faint stars in the Solar neighborhood, and faint stars
with extreme transverse velocity components (possibly halo stars). Our
spectroscopic follow-up survey of new high proper motion stars is
still in progress, and is being expanded as new northern stars with
large proper motions are being discovered. Future results will be
presented in upcoming papers in this series.

\acknowledgments

This research program is being supported by NSF grant AST-0087313 at
the American Museum of Natural History, as part of the NStars program.

This work has been made possible through the use of the Guide Star
Catalogue-II. The Guide Star Catalogue-II is a joint project of the
Space Telescope Science Institute and the Osservatorio Astronomico di
Torino. Space Telescope Science Institute is operated by the
Association of Universities for Research in Astronomy, for the
National Aeronautics and Space Administration under contract
NAS5-26555. The participation of the Osservatorio Astronomico di
Torino is supported by the Italian Council for Research in
Astronomy. Additional support is provided by European Southern
Observatory, Space Telescope European Coordinating Facility, the
International GEMINI project and the European Space Agency
Astrophysics Division.

This publication makes use of data products from the Two Micron All
Sky Survey, which is a joint project of the University of
Massachusetts and the Infrared Processing and Analysis Center, funded
by the National Aeronautics and Space Administration and the National
Science Foundation.

The data mining required for this work has been made possible with the
use of the SIMBAD astronomical database and VIZIER astronomical
catalogs service, both maintained and operated by the Centre de
Donn\'ees Astronomiques de Strasbourg (http://cdsweb.u-strasbg.fr/).

We thank the anonymous referee for a very thorough review, which lead
to a significant improvement of the manuscript.


\begin{deluxetable}{lrrrrrrrrr}
\tabletypesize{\scriptsize}
\tablecolumns{10} 
\tablewidth{0pt} 
\tablecaption{Newly classified high proper motion stars} 
\tablehead{
\colhead{Star} & 
\colhead{$\mu$($\arcsec$ yr$^{-1}$)} & 
\colhead{pma ($\deg$)} & 
\colhead{B} &
\colhead{R} &
\colhead{${\rm K_s}$} &
\colhead{Date of spec.} &
\colhead{Obs.\tablenotemark{1}} &
\colhead{Sp. type\tablenotemark{2}} &
\colhead{HRV(km s$^{-1}$)\tablenotemark{3}}
}
\startdata 
LSR0011+5908  &  1.483& 218.3&\nodata& 14.5&\nodata& 2001-07-24& L & M5.5 V  & +90$\pm$40 \\
LSR0014+6546  &  0.962&  67.2& 18.0& 15.7&\nodata& 2001-07-24& L & sdM4.5  &-150$\pm$40 \\
LSR0020+5526  &  0.541&  55.7& 17.2& 14.9&\nodata& 2001-02-03& M & esdM2.5 &-100$\pm$50 \\
LSR0124+6819  &  0.559& 131.4& 18.9& 16.4&\nodata& 2001-12-08& L & M7.0 V  & -10$\pm$30 \\
LSR0134+6459  &  0.924&  75.5& 17.9& 15.2&\nodata& 2001-12-08& L & M5.5 V  &  +0$\pm$30 \\
LSR0155+3758  &  0.538& 153.9& 17.0& 14.8&  9.6& 2001-07-24& L & M5.0 V  & -60$\pm$40 \\
LSR0157+5308  &  0.641&  94.7& 17.1& 14.9& 11.3& 2001-02-02& M & sdM3.5  & -30$\pm$30 \\
LSR0200+5530  &  0.507& 112.0&\nodata& 19.7&\nodata& 2001-02-05& M & M5.5 V  & -70$\pm$50 \\
LSR0212+7012  &  0.744&  76.2& 18.4& 15.9&\nodata& 2001-12-08& L & M5.0 V  & +10$\pm$30 \\
LSR0258+5354  &  0.545& 127.7& 17.2& 14.9& 12.6& 2001-02-02& M & sdK7    &+130$\pm$30 \\
LSR0310+6634  &  0.808& 121.6&\nodata& 17.5&\nodata& 2001-07-25& L & DC10    &     \nodata \\
LSR0316+3132  &  0.759& 139.2& 17.7& 15.4& 10.6& 2001-12-08& L & M5.0 V  &+100$\pm$30 \\
LSR0340+5124  &  0.932& 155.9& 17.6& 15.4&\nodata& 2001-02-01& M & M5.5 V  & -30$\pm$30 \\
LSR0342+5527  &  0.500& 145.8& 17.0& 14.8&\nodata& 2001-02-04& M & sdM0.0  & +90$\pm$30 \\
LSR0354+3333  &  0.848& 150.2&\nodata& 16.4& 10.9& 2001-02-02& M & M6.0 V  & +70$\pm$30 \\
LSR0358+8111  &  0.547& 143.8& 19.0& 16.9& 13.6& 2001-12-08& L & sdM1.5  & -30$\pm$30 \\
LSR0400+5417  &  0.754& 136.7& 18.0& 16.0&\nodata& 2001-02-04& M & (e)sdK7 &-150$\pm$30 \\
LSR0401+5131  &  0.886& 156.0& 17.9& 16.7&\nodata& 2001-02-04& M & DC8     &     \nodata \\
LSR0419+4233  &  1.535& 159.4&\nodata& 17.4&\nodata& 2001-02-02& M & M8.5 Ve & +30$\pm$30 \\
LSR0455+0244  &  0.768& 123.2& 18.5& 16.4&\nodata& 2001-12-08& L & M5.5 V  & +60$\pm$30 \\
LSR0455+5252  &  0.804& 187.8&\nodata& 18.3&\nodata& 2001-02-02& M & M8.0 Ve & -30$\pm$30 \\
LSR0505+3043  &  1.097& 148.5& 18.5& 16.0&\nodata& 2001-02-01& M & esdM3.5 &+210$\pm$30 \\
LSR0505+6633  &  0.577& 155.1&\nodata& 17.0&\nodata& 2001-07-26& L & sdM4.5  & -90$\pm$40 \\
LSR0515+5911  &  1.015& 173.3&\nodata& 16.8&\nodata& 2001-07-26& L & M7.0 V  & -60$\pm$40 \\
LSR0519+4213  &  1.181& 139.5&\nodata& 16.1& 12.9& 2001-02-04& M & esdM3.5 &-280$\pm$30 \\
LSR0521+3425  &  0.512& 149.2& 18.1& 15.8& 11.0& 2001-02-04& M & M5.0 V  & +60$\pm$30 \\
LSR0522+3814  &  1.703& 164.1& 16.6& 14.5& 12.6& 2001-02-01& M & esdM3.0 &-110$\pm$30 \\
LSR0524+3358  &  0.530& 138.4&\nodata& 17.5& 14.1& 2001-02-04& M & sdM1.5  &-190$\pm$50 \\
LSR0527+3009  &  0.639& 221.0&\nodata& 16.3& 11.5& 2001-02-05& M & M5.0 V  &  +0$\pm$30 \\
LSR0533+3837  &  0.551& 134.9& 18.1& 16.3& 13.1& 2001-02-05& M & sdM2.0  &-170$\pm$30 \\
LSR0539+4038  &  1.057& 141.8&\nodata& 17.0& 10.0& 2001-02-01& M & M8.0 Ve & +10$\pm$30 \\
LSR0541+3959  &  0.566& 143.2& 18.1& 17.1& 15.3& 2001-02-05& M & DA8     & +10$\pm$30 \\
LSR0549+2329  &  1.379& 131.4&\nodata& 17.4&\nodata& 2001-02-02& M & DC8     &     \nodata \\
LSR0556+1144  &  0.611& 121.8& 16.5& 14.2&\nodata& 2001-12-08& L & M5.5 V  &+160$\pm$30 \\
LSR0609+2319  &  1.104& 130.7& 18.7& 16.5& 12.4& 2001-12-08& L & sdM5.0  & -40$\pm$30 \\
LSR0618+1614  &  0.646& 165.7& 16.9& 14.7& 11.9& 2001-02-04& M & sdM2.0  & +70$\pm$30 \\
LSR0621+3652  &  0.865& 176.9& 16.6& 14.5& 11.9& 2001-02-02& M & esdK7   &+230$\pm$30 \\
LSR0627+0616  &  1.019& 178.5& 17.6& 15.4&\nodata& 2001-02-01& M & esdM1.5 &-130$\pm$30 \\
LSR0628+0529  &  0.551& 140.4&\nodata& 17.1&\nodata& 2001-02-05& M & M7.0 V  &+110$\pm$30 \\
LSR0646+3212  &  0.567& 142.4&\nodata& 17.5& 13.1& 2001-02-05& M & M5.5 V  &+170$\pm$80 \\
LSR0702+2154  &  0.653& 144.7& 16.9& 14.9& 10.3& 2001-02-02& M & M5.5 V  & +20$\pm$30 \\  
LSR0705+0506  &  0.510& 158.4& 18.8& 16.0&\nodata& 2001-02-04& M & sdM3.5  &+100$\pm$50 \\
LSR0721+3714  &  0.587& 234.7& 18.9& 16.2& 10.8& 2001-02-02& M & M5.5 Ve & +60$\pm$30 \\
LSR0731+0729  &  0.512& 190.3& 17.3& 15.2&\nodata& 2001-02-02& M & M5.0 V  &+150$\pm$50 \\
LSR0803+1547  &  0.516& 154.3& 18.4& 16.1& 13.4& 2001-02-02& M & sdM0.0  &-140$\pm$30 \\
LSR1722+1004  &  0.724& 192.1& 17.5& 15.1&\nodata& 2001-07-24& L & M4.0 V  & -10$\pm$40 \\
LSR1755+1648  &  0.995& 116.9& 16.4& 14.2&\nodata& 2001-07-24& L & sdM3.5  & -30$\pm$40 \\
LSR1757+0015  &  0.518& 184.8& 19.4& 17.0&\nodata& 2001-07-24& L & sdM4.5  & +10$\pm$40 \\
LSR1758+1417  &  1.014& 235.4& 16.9& 15.8& 14.6& 2001-07-24& L & DA10    &     \nodata \\
LSR1802+0028  &  0.543& 211.3&\nodata& 17.6&\nodata& 2001-07-26& L & esdM1.5 &-480$\pm$40 \\
LSR1806+1141  &  0.541& 186.0& 16.5& 14.4&\nodata& 2001-07-26& L & M4.0 V  & +10$\pm$40 \\
LSR1808+1134  &  0.606& 228.7& 16.8& 14.4&\nodata& 2001-07-26& L & M5.0 V  & +10$\pm$40 \\
LSR1809-0219  &  0.506& 176.0& 16.4& 13.9&  9.3& 2001-07-25& L & M4.5 V  & +40$\pm$40 \\
LSR1809-0247  &  1.005& 214.9& 17.4& 15.2& 10.7& 2001-07-25& L & M5.0 V  & -80$\pm$40 \\
LSR1817+1328  &  1.207& 201.5& 16.7& 15.5&\nodata& 2001-07-25& L & DA10    & -40$\pm$40 \\
LSR1820-0031  &  0.555& 199.3& 17.8& 15.7&\nodata& 2001-07-26& L & sdM2.0  & -10$\pm$40 \\
LSR1833+2219  &  0.502& 200.0& 17.6& 14.9&\nodata& 2001-07-24& L & M5.0 V  & +40$\pm$30 \\
LSR1835+3259  &  0.747& 185.8&\nodata& 16.6&  9.2& 2002-05-25& M & M9.0 Ve & -10$\pm$40 \\
LSR1836+1040  &  0.921& 206.0& 18.2& 15.9&\nodata& 2001-07-25& L & esdM0.5 &-150$\pm$40 \\
LSR1841+2421  &  0.752& 189.0& 19.0& 16.5&\nodata& 2001-02-03& M & M6.0 V  & +30$\pm$30 \\
LSR1843+0507  &  0.577& 253.2& 19.2& 17.1&\nodata& 2001-07-26& L & M5.5 V  & -50$\pm$40 \\
LSR1844+0947  &  0.501& 224.3& 17.6& 15.8&\nodata& 2001-07-26& L &(sd)M2.0V& -90$\pm$40 \\
LSR1851+2641  &  0.704&  26.3& 18.9& 16.1&\nodata& 2001-02-03& M & M6.0 Ve & +10$\pm$30 \\
LSR1859+0156  &  0.674& 142.2& 16.9& 14.7&\nodata& 2001-07-26& L & M4.5 V  & -90$\pm$40 \\
LSR1914+2825A &  0.529& 212.6& 17.5& 15.0&\nodata& 2001-07-26& L & sdM0.0  & -80$\pm$40 \\
LSR1914+2825B &  0.531& 214.3& 18.7& 15.6&\nodata& 2001-07-26& L & sdM1.5  &-120$\pm$40 \\
LSR1918+1728  &  0.626& 192.2&\nodata& 17.2&\nodata& 2001-07-25& L & esdM3.0 &-160$\pm$40 \\
LSR1919+1438  &  0.507&  96.6&\nodata& 14.9&\nodata& 2001-07-26& L & M5.0 V  & -20$\pm$40 \\
LSR1922+4605  &  0.555& 206.3& 17.8& 15.1& 12.9& 2002-05-25& M & sdM0.0  & -20$\pm$40 \\
LSR1927+6802  &  0.515& 227.2&\nodata& 17.2&\nodata& 2002-05-25& M & M6.5 V  & -20$\pm$40 \\
LSR1928-0200A &  0.858& 194.7&\nodata& 15.2& 11.3& 2001-07-25& L & M3.5 V  &-100$\pm$40 \\
LSR1928-0200B &  0.858& 194.7&\nodata& 18.2& 13.1& 2001-07-25& L & M5.5 V  & -80$\pm$40 \\
LSR1933-0138  &  0.895& 132.6& 15.6& 13.5& 10.1& 2001-07-26& L & M3.0 V  &-170$\pm$40 \\
LSR1943+0941  &  0.543& 213.6& 18.7& 16.2&\nodata& 2002-05-25& M & M5.5 V  &  +0$\pm$40 \\
LSR1945+4650A &  0.612& 228.0& 17.9& 16.8& 15.5& 2002-05-25& M & DA9     &     \nodata \\
LSR1945+4650B &  0.609& 228.7& 19.3& 17.0& 15.2& 2002-05-25& M & DC10    &     \nodata \\
LSR1946+0937  &  0.566& 173.7& 17.5& 16.8&\nodata& 2001-07-26& L & DA9     &-140$\pm$40 \\
LSR1946+0942  &  0.555& 173.8& 16.8& 14.6&\nodata& 2001-07-26& L & M3.5 V  &-130$\pm$40 \\
LSR1956+4428  &  0.891& 214.3& 17.3& 15.0&\nodata& 2000-08-01& L & esdM0.5 &-140$\pm$30 \\
LSR2000+0404  &  0.503& 125.1& 18.0& 15.7&\nodata& 2001-07-26& L & M5.5 V  & +20$\pm$40 \\
LSR2000+3057  &  1.339&  16.7&\nodata& 15.6&  9.7& 2001-07-24& L & M6.0 Ve & -10$\pm$40 \\
LSR2005+0835  &  0.583& 197.9& 16.5& 14.6&\nodata& 2001-07-26& L & sdK5    & -20$\pm$40 \\
LSR2009+5659  &  0.824&  31.5& 16.3& 14.2& 11.2& 2001-07-25& L & sdM2.0  &-120$\pm$40 \\
LSR2010+3938  &  0.512& 180.4& 17.4& 13.3& 11.8& 2000-08-01& L &(sd)M1.5V&-210$\pm$30 \\
LSR2013+0417  &  0.749& 195.5& 16.2& 14.4&\nodata& 2001-07-25& L & sdK7    &-190$\pm$40 \\
LSR2017+0623  &  0.674& 187.3& 18.2& 16.0&\nodata& 2001-07-25& L & M5.0 V  & +20$\pm$40 \\
LSR2036+5059  &  1.054& 100.8&\nodata& 16.8&\nodata& 2001-12-09& L & sdM7.5  &-140$\pm$30 \\
LSR2044+1339  &  0.518&  31.2& 15.4& 15.4& 10.3& 2001-07-26& L & M5.0 Ve & -20$\pm$40 \\
LSR2050+7740  &  0.543&  27.3& 19.3& 18.0&\nodata& 2002-05-20& K & DC11    &     \nodata \\
LSR2059+5517  &  0.500&  33.1& 18.2& 16.8&\nodata& 2002-05-25& M & DC11    &     \nodata \\
LSR2107+3600  &  0.736& 231.4& 18.4& 15.9& 11.7& 2001-07-25& L & M4.5 V  &-130$\pm$40 \\
LSR2115+3804  &  0.506&  25.2& 17.3& 15.0& 12.3& 2001-07-26& L & esdK7   &-140$\pm$40 \\
LSR2117+7345  &  0.746&  41.1& 18.6& 16.4&\nodata& 2001-07-24& L & M6.0 V  & +70$\pm$40 \\
LSR2122+3656  &  0.816&  65.0& 18.7& 16.2&\nodata& 2001-07-24& L & esdM5.0 &-100$\pm$40 \\
LSR2124+4003  &  0.697&  50.9& 17.3& 14.9&\nodata& 2000-08-02& L & M6.5 V  & -60$\pm$30 \\
LSR2132+4754  &  0.569&  37.8& 16.8& 14.5& 10.7& 2001-07-26& L & M4.0 V  & -60$\pm$40 \\
LSR2146+5147  &  0.584&  49.7& 18.5& 16.0&\nodata& 2001-07-26& L & sdM1.0  &-210$\pm$40 \\
LSR2158+6117  &  0.819&  82.3& 17.9& 15.7&\nodata& 2001-07-24& L & M6.0 V  & -70$\pm$40 \\
LSR2205+5353  &  0.528& 236.4& 17.3& 15.1&\nodata& 2001-07-25& L & sdM1.0  &-200$\pm$40 \\
LSR2205+5807  &  0.538& 108.9& 17.9& 17.6&\nodata& 2001-07-26& L & esdM1.0 &-130$\pm$40 \\
LSR2251+4706  &  0.631&  71.5&\nodata& 17.9& 12.6& 2001-07-25& L & M6.5 V  &-100$\pm$40 \\
LSR2311+5032  &  0.669&  81.0& 16.9& 14.6&  9.9& 2001-02-03& M & M4.5 V  & -30$\pm$30 \\
LSR2311+5103  &  0.531& 253.5&\nodata& 18.1& 12.3& 2001-07-25& L & M7.5 V  & -10$\pm$40 \\
LSR2321+4704  &  0.712&  70.5& 18.5& 16.2& 13.2& 2001-12-08& L & esdM2.0 &-220$\pm$30 
\enddata
\tablenotetext{1}{Observatory used: Lick 3m Shane (L), MDM 2.4m Hiltner (M),
 KPNO 4m Mayall (K)}
\tablenotetext{2}{Spectral class and subtype based on
spectroscopy. Parenthesis indicates possible change in spectral class
required for consistency with the star's kinematics.}
\tablenotetext{3}{Estimated heliocentric radial velocity.}
\end{deluxetable} 

\clearpage

\begin{deluxetable}{ccccc}
\tabletypesize{\scriptsize}
\tablecolumns{7} 
\tablewidth{0pt} 
\tablecaption{Spectral Type Indices for M dwarfs and subdwarfs}
\tablehead{
\colhead{Index Name} & 
\colhead{Numerator} &
\colhead{Denominator} &
\colhead{Other Name} & 
\colhead{Ref.}
}
\startdata 
CaH1   & 6380-6390 & Avg. of 6410-6420 and 6345-6355& \nodata  & Reid {\it et al.} 1995 \\
CaH2   & 6814-6846 & 7042-7046 & \nodata  & Reid {\it et al.} 1995 \\
CaH3   & 6960-6990 & 7042-7046 & \nodata  & Reid {\it et al.} 1995 \\
TiO5   & 7126-7135 & 7042-7046 & \nodata  & Reid {\it et al.} 1995 \\
VO1    & 7430-7470 & 7550-7570 & VO 7434  & Hawley {\it et al.} 2002 \\
TiO6   & 7550-7570 & 7745-7765 & \nodata  & this paper \\
VO2    & 7920-7960 & 8130-8150 & \nodata  & this paper \\
TiO7   & 8440-8470 & 8400-8420 & TiO 8440 & Hawley {\it et al.} 2002 \\
Color-M & 8105-8155 & 6510-6560 & \nodata  & this paper
\enddata     
\end{deluxetable} 

\begin{deluxetable}{lrrrrrrrrrc}
\tabletypesize{\scriptsize}
\tablecolumns{11} 
\tablewidth{0pt} 
\tablecaption{Measured Spectral Indices of observed M dwarfs,
subdwarfs, and extreme subdwarfs}
\tablehead{
\colhead{Star} & 
\colhead{CaH1} &
\colhead{CaH2} & 
\colhead{CaH3} &
\colhead{TiO5} &
\colhead{VO1} &
\colhead{TiO6} &
\colhead{VO2} &
\colhead{TiO7} &
\colhead{Color-M} &
\colhead{Sp. Type}
}
\startdata 
LSR0011+5908   & 0.774& 0.266& 0.556& 0.214& 0.843& 0.548& 0.641& 0.739&  3.632& M5.5V   \\
LSR0014+6546   & 0.574& 0.369& 0.575& 0.628& 0.926& 0.984& 0.954& 0.994&  1.479& sdM4.5  \\
LSR0020+5526   & 0.566& 0.471& 0.654& 0.986& 0.964& 1.025& 0.896& 1.070&  1.568& esdM2.5 \\
LSR0124+6819   & 0.928& 0.294& 0.625& 0.223& 0.776& 0.483& 0.489& 0.595&  5.519& M7.0V   \\
LSR0134+6459   & 0.848& 0.340& 0.664& 0.270& 0.874& 0.580& 0.631& 0.742&  3.600& M5.5V   \\
LSR0155+3758   & 0.823& 0.378& 0.685& 0.326& 0.861& 0.581& 0.682& 0.802&  2.833& M5.0V   \\
LSR0157+5308   & 0.681& 0.426& 0.630& 0.505& 0.975& 0.827& 0.915& 0.913&  1.200& sdM3.5  \\
LSR0200+5530   & 0.498& 0.254& 0.538& 0.238& 0.743& 0.495& 0.842& 0.811&  4.247& M5.5V   \\
LSR0212+7012   & 0.673& 0.224& 0.570& 0.201& 0.818& 0.663& 0.730& 0.795&  3.721& M5.0V   \\
LSR0258+5354   & 0.898& 0.844& 0.941& 0.958& 0.967& 0.949& 0.983& 0.916&  1.053& sdK7    \\
LSR0316+3132   & 0.799& 0.364& 0.647& 0.293& 0.883& 0.656& 0.696& 0.799&  3.023& M5.0V   \\
LSR0340+5124   & 0.831& 0.349& 0.629& 0.326& 0.900& 0.609& 0.683& 0.603&  2.918& M5.5V   \\
LSR0342+5527   & 0.803& 0.733& 0.866& 0.925& 0.983& 0.976& 1.028& 1.073&  0.980& sdM0.0  \\
LSR0354+3333   & 0.829& 0.305& 0.652& 0.236& 0.854& 0.492& 0.555& 0.635&  3.317& M6.0V   \\
LSR0358+8111   & 0.660& 0.571& 0.795& 0.802& 0.991& 0.989& 1.009& 1.013&  1.355& sdM1.5  \\
LSR0400+5417   & 0.879& 0.776& 0.868& 0.884& 0.969& 0.944& 1.112& 1.715&  0.947& (e)sdK7 \\
LSR0419+4233   & 0.923& 0.362& 0.716& 0.315& 0.762& 0.457& 0.349& 0.467&  7.267& M8.5Ve  \\
LSR0455+0244   & 0.672& 0.252& 0.539& 0.284& 0.860& 0.572& 0.648& 0.776&  3.636& M5.5V   \\
LSR0455+5252   & 0.750& 0.238& 0.517& 0.112& 0.811& 0.321& 0.417& 0.426&  5.076& M8.0Ve  \\
LSR0505+3043   & 0.616& 0.397& 0.626& 0.797& 0.961& 0.901& 0.887& 0.799&  1.316& esdM3.5 \\
LSR0505+6633   & 0.500& 0.340& 0.584& 0.591& 1.017& 1.071& 0.955& 1.006&  1.426& sdM4.5  \\
LSR0515+5911   & 0.808& 0.257& 0.563& 0.209& 0.814& 0.453& 0.471& 0.612&  6.112& M7.0V   \\
LSR0519+4213   & 0.659& 0.414& 0.583& 0.785& 0.951& 0.993& 1.045& 0.980&  1.124& esdM3.5 \\
LSR0521+3425   & 0.828& 0.380& 0.691& 0.303& 0.885& 0.579& 0.721& 0.776&  2.245& M5.0V   \\
LSR0522+3814   & 0.566& 0.462& 0.642& 0.970& 0.965& 1.027& 0.900& 1.064&  1.588& esdM3.0 \\
LSR0524+3358   & 0.945& 0.589& 0.787& 0.810& 0.924& 0.916& 0.967& 1.050&  1.247& sdM1.5  \\
LSR0527+3009   & 0.778& 0.295& 0.591& 0.266& 0.928& 0.633& 0.690& 0.801&  2.669& M5.0V   \\
LSR0533+3837   & 0.690& 0.533& 0.758& 0.776& 0.950& 0.942& 1.001& 0.905&  1.362& sdM2.0  \\
LSR0539+4038   & 0.854& 0.216& 0.544& 0.158& 0.846& 0.371& 0.395& 0.436&  5.494& M8.0Ve  \\
LSR0556+1144   & 0.772& 0.324& 0.649& 0.252& 0.850& 0.546& 0.600& 0.712&  3.864& M5.5V   \\
LSR0609+2319   & 0.660& 0.322& 0.553& 0.364& 0.911& 0.800& 0.835& 0.899&  2.188& sdM5.0  \\
LSR0618+1614   & 0.724& 0.534& 0.760& 0.812& 1.010& 0.998& 1.006& 0.840&  1.176& sdM2.0  \\
LSR0621+3652   & 0.827& 0.768& 0.896& 0.981& 1.025& 1.017& 0.989& 1.047&  1.104& esdK7   \\
LSR0627+0616   & 0.683& 0.583& 0.772& 0.950& 0.998& 0.952& 0.887& 1.017&  1.327& esdM1.5 \\
LSR0628+0529   & 0.836& 0.228& 0.583& 0.210& 0.837& 0.473& 0.514& 0.595&  5.167& M7.0V   \\
LSR0646+3212   & 0.596& 0.312& 0.555& 0.171& 0.838& 0.708& 0.661& 1.135&  5.019& M5.5V   \\
LSR0702+2154   & 0.801& 0.353& 0.680& 0.307& 0.890& 0.542& 0.630& 0.771&  2.530& M5.5V   \\
LSR0705+0506   & 0.688& 0.399& 0.648& 0.536& 0.974& 0.832& 0.914& 0.863&  1.255& sdM3.5  \\
LSR0721+3714   & 0.803& 0.295& 0.593& 0.254& 0.856& 0.530& 0.628& 0.716&  3.533& M5.5Ve  \\
LSR0731+0729   & 0.832& 0.346& 0.674& 0.301& 0.907& 0.581& 0.697& 0.741&  2.780& M5.0V   \\
LSR0803+1548   & 0.781& 0.723& 0.818& 0.927& 1.005& 0.913& 1.080& 0.882&  1.330& sdM0.0  \\
LSR1722+1004   & 0.789& 0.403& 0.679& 0.397& 0.932& 0.710& 0.814& 0.897&  2.029& M4.0V   \\
LSR1755+1648   & 0.707& 0.419& 0.664& 0.486& 0.943& 0.814& 0.886& 0.953&  1.691& sdM3.5  \\
LSR1757+0015   & 0.666& 0.356& 0.599& 0.429& 0.935& 0.783& 0.835& 0.951&  1.947& sdM4.5  \\
LSR1802+0028   & 0.833& 0.570& 0.757& 0.998& 0.981& 1.040& 0.967& 1.082&  1.295& esdM1.5 \\
LSR1806+1141   & 0.846& 0.435& 0.718& 0.392& 0.926& 0.666& 0.751& 0.862&  2.307& M4.0V   \\
LSR1808+1134   & 0.907& 0.409& 0.720& 0.330& 0.894& 0.586& 0.658& 0.780&  2.943& M5.0V   \\
LSR1809-0219   & 0.816& 0.364& 0.644& 0.330& 0.904& 0.643& 0.713& 0.805&  2.684& M4.5V   \\
LSR1809-0247   & 0.765& 0.313& 0.594& 0.272& 0.918& 0.624& 0.722& 0.827&  2.639& M5.0V   \\
LSR1820-0031   & 0.737& 0.518& 0.711& 0.658& 1.010& 0.901& 0.933& 1.017&  1.440& sdM2.0  \\
LSR1833+2219B  & 0.827& 0.362& 0.662& 0.304& 0.885& 0.595& 0.688& 0.792&  2.711& M5.0V   \\
LSR1835+3259   & 0.806& 0.279& 0.547& 0.277& 0.812& 0.454& 0.323& 0.452& 13.194& M9.0Ve  \\
LSR1836+1040   & 0.754& 0.678& 0.812& 0.990& 0.984& 1.042& 0.981& 1.050&  1.176& esdM0.5 \\
LSR1841+2421   & 0.741& 0.245& 0.565& 0.230& 0.862& 0.519& 0.653& 0.611&  4.387& M6.0V   \\
LSR1843+0507   & 0.813& 0.303& 0.587& 0.251& 0.891& 0.536& 0.625& 0.752&  3.412& M5.5V   \\
LSR1844+0947   & 0.802& 0.515& 0.749& 0.547& 0.969& 0.830& 0.873& 0.965&  1.600& (sd)M2.0V   \\
LSR1851+2641   & 0.758& 0.235& 0.515& 0.200& 0.882& 0.564& 0.646& 0.619&  4.177& M6.0Ve  \\
LSR1859+0156   & 0.816& 0.375& 0.664& 0.318& 0.912& 0.603& 0.697& 0.841&  2.651& M4.5V   \\
LSR1914+2825A  & 0.821& 0.691& 0.844& 0.821& 0.953& 0.939& 0.954& 1.015&  1.260& sdM0.0  \\
LSR1914+2825B  & 0.766& 0.560& 0.748& 0.742& 0.987& 0.941& 0.936& 1.014&  1.382& sdM1.5  \\
LSR1918+1728   & 0.647& 0.439& 0.666& 0.830& 0.968& 1.034& 0.986& 0.979&  1.340& esdM3.0 \\
LSR1919+1438   & 0.808& 0.364& 0.661& 0.307& 0.901& 0.596& 0.671& 0.797&  2.978& M5.0V   \\
LSR1922+4605   & 0.795& 0.729& 0.835& 0.854& 0.983& 1.011& 0.970& 0.994&  1.259& sdM0.0  \\
LSR1927+6802   & 0.746& 0.326& 0.688& 0.225& 0.852& 0.510& 0.546& 0.662&  4.693& M6.5V   \\
LSR1928-0200A  & 0.851& 0.474& 0.735& 0.460& 0.940& 0.719& 0.844& 0.939&  1.773& M3.5V   \\
LSR1928-0200B  & 0.840& 0.295& 0.602& 0.235& 0.887& 0.539& 0.630& 0.768&  3.340& M5.5V   \\
LSR1933-0138   & 0.818& 0.476& 0.735& 0.480& 0.952& 0.747& 0.849& 0.988&  1.743& M3.0V   \\
LSR1943+0941   & 0.908& 0.336& 0.652& 0.269& 0.867& 0.570& 0.598& 0.733&  3.673& M5.5V   \\
LSR1946+0942   & 0.732& 0.423& 0.676& 0.478& 0.938& 0.780& 0.852& 0.960&  1.763& M3.5V   \\
LSR1956+4428   & 0.000& 0.679& 0.820& 1.011& 0.979& 1.034& 0.960& 1.035&  1.303& esdM0.5 \\
LSR2000+0404   & 0.850& 0.308& 0.610& 0.243& 0.875& 0.511& 0.604& 0.751&  3.589& M5.5V   \\
LSR2000+3057   & 0.770& 0.246& 0.521& 0.198& 0.855& 0.514& 0.613& 0.708&  4.064& M6.0Ve  \\
LSR2005+0835   & 0.959& 0.954& 0.978& 1.025& 0.991& 0.969& 0.981& 1.019&  1.054& sdK5    \\
LSR2009+5659   & 0.678& 0.519& 0.718& 0.758& 0.967& 0.981& 0.974& 1.017&  1.304& sdM2.0  \\
LSR2010+3938   & 0.000& 0.552& 0.785& 0.667& 0.971& 0.917& 0.882& 1.002&  1.562& (sd)M1.5V\\
LSR2013+0417   & 0.924& 0.893& 0.927& 0.997& 0.997& 0.988& 1.002& 1.023&  1.042& sdK7    \\
LSR2017+0623   & 0.800& 0.311& 0.597& 0.288& 0.896& 0.640& 0.736& 0.837&  2.683& M5.0V   \\
LSR2036+5059   & 0.572& 0.173& 0.366& 0.263& 0.867& 0.816& 0.841& 0.894&  2.788& sdM7.5  \\
LSR2044+1339   & 0.811& 0.349& 0.629& 0.314& 0.886& 0.593& 0.695& 0.810&  2.868& M5.0Ve  \\
LSR2107+3600   & 0.735& 0.353& 0.620& 0.350& 0.892& 0.688& 0.779& 0.863&  2.271& M4.5V   \\
LSR2115+3804   & 0.852& 0.798& 0.900& 1.019& 0.959& 0.984& 0.987& 1.022&  1.113& esdK7   \\
LSR2117+7345   & 0.798& 0.289& 0.610& 0.214& 0.849& 0.516& 0.590& 0.717&  3.624& M6.0V   \\
LSR2122+3656   & 0.430& 0.304& 0.463& 0.807& 0.951& 1.055& 0.999& 1.026&  1.415& esdM5.0 \\
LSR2124+4003   & 0.000& 0.359& 0.710& 0.282& 0.841& 0.561& 0.575& 0.707&  4.122& M6.5V   \\
LSR2132+4754   & 0.831& 0.413& 0.690& 0.412& 0.909& 0.684& 0.786& 0.892&  2.133& M4.0V   \\
LSR2146+5147   & 0.752& 0.603& 0.777& 0.860& 0.964& 0.928& 0.928& 1.032&  1.240& sdM1.0  \\
LSR2158+6117   & 0.779& 0.269& 0.575& 0.209& 0.845& 0.543& 0.616& 0.730&  3.610& M6.0V   \\
LSR2205+5353   & 0.721& 0.612& 0.782& 0.950& 0.991& 1.019& 0.973& 1.029&  1.199& esdM1.0 \\
LSR2205+5807   & 0.780& 0.630& 0.745& 0.867& 0.983& 1.025& 0.920& 1.058&  1.607& sdM1.0  \\
LSR2251+4706   & 0.740& 0.229& 0.535& 0.154& 0.835& 0.489& 0.568& 0.662&  4.601& M6.5V   \\
LSR2311+5032   & 0.836& 0.366& 0.666& 0.335& 0.898& 0.617& 0.750& 0.792&  2.850& M4.5V   \\
LSR2311+5103   & 1.064& 0.258& 0.628& 0.160& 0.811& 0.463& 0.468& 0.604&  6.292& M7.5V 	 \\
LSR2321+4704   & 0.797& 0.535& 0.720& 0.858& 0.945& 1.043& 0.973& 0.993&  1.323& esdM2.0
\enddata
\end{deluxetable} 

\clearpage

\begin{deluxetable}{lrrrrr}
\tabletypesize{\scriptsize}
\tablecolumns{6} 
\tablewidth{0pt} 
\tablecaption{Estimated distances and kinematics} 
\tablehead{
\colhead{Star} & 
\colhead{Sp. type\tablenotemark{1}} &
\colhead{d(pc)\tablenotemark{2}} &
\colhead{U (km s$^{-1}$)\tablenotemark{3}} &
\colhead{V (km s$^{-1}$)\tablenotemark{3}} &
\colhead{W (km s$^{-1}$)\tablenotemark{3}}
}
\startdata 
LSR0011+5908  &   M5.5 V  &  12$\pm$4     &   0$\pm$30     & 100$\pm$40      & -70$\pm$20 \\
LSR0014+6546  &   sdM4.5  &  40$\pm$20    & -90$\pm$80     &-220$\pm$60      &  30$\pm$20 \\
LSR0020+5526  &   esdM2.5 &  70$\pm$35    &-120$\pm$70     &-160$\pm$60      &  90$\pm$40 \\
LSR0124+6819  &   M7.0 V  &  11$\pm$3     & -20$\pm$20     & -30$\pm$20      & -20$\pm$10 \\
LSR0134+6459  &   M5.5 V  &  18$\pm$6     & -70$\pm$30     & -50$\pm$30      &  30$\pm$10 \\
LSR0155+3758  &   M5.0 V  &  20$\pm$6     &  20$\pm$30     & -80$\pm$30      & -20$\pm$20 \\
LSR0157+5308  &   sdM3.5  &  45$\pm$22    & -90$\pm$50     &-120$\pm$50      &  20$\pm$10 \\
LSR0200+5530  &   M5.5 V  &  90$\pm$30    &-120$\pm$60     &-200$\pm$60      & -20$\pm$10 \\
LSR0212+7012  &   M5.0 V  &  40$\pm$13    &-100$\pm$40     & -80$\pm$30      &  70$\pm$20 \\
LSR0258+5354  &   sdK7    & 250$\pm$100   &-500$\pm$160    &-430$\pm$200     &-120$\pm$40 \\
LSR0310+6634  &   DC10    &  60$\pm$30    &-170$\pm$80     &-170$\pm$80      & -10$\pm$10 \\
LSR0316+3132  &   M5.0 V  &  30$\pm$10    &-130$\pm$30     & -70$\pm$30      & -70$\pm$10 \\
LSR0340+5124  &   M5.5 V  &  18$\pm$6     & -20$\pm$30     & -80$\pm$30      & -40$\pm$10 \\
LSR0342+5527  &   sdM0.0  & 150$\pm$75    &-270$\pm$100    &-230$\pm$140     &-120$\pm$60 \\
LSR0354+3333  &   M6.0 V  &  20$\pm$6     & -90$\pm$30     & -60$\pm$20      & -50$\pm$10 \\
LSR0358+8111  &   sdM1.5  & 200$\pm$100   &-400$\pm$200    &-330$\pm$170     &-100$\pm$40 \\
LSR0400+5417  &   (e)sdK7 & 375$\pm$180   &-580$\pm$340   &-1220$\pm$550     &-140$\pm$60 \\
\nodata       &   \nodata &(150$\pm$75)  &(-160$\pm$140)   &(-540$\pm$230)   &(-60$\pm$30)\\
LSR0401+5131  &   DC8     &  26$\pm$13    & -60$\pm$20     & -90$\pm$40      & -50$\pm$20 \\
LSR0419+4233  &   M8.5 Ve &  10$\pm$3     & -60$\pm$30     & -60$\pm$20      & -40$\pm$10 \\
LSR0455+0244  &   M5.5 V  &  28$\pm$9     & -50$\pm$30     &-110$\pm$30      &  10$\pm$20 \\
LSR0455+5252  &   M8.0 Ve &  20$\pm$6     & -10$\pm$30     & -60$\pm$20      & -60$\pm$20 \\
LSR0505+3043  &   esdM3.5 &  70$\pm$35    &-250$\pm$40     &-340$\pm$180     & -60$\pm$20 \\
LSR0505+6633  &   sdM4.5  &  70$\pm$35    & -50$\pm$60     &-200$\pm$80      & -60$\pm$20 \\
LSR0515+5911  &   M7.0 V  &  14$\pm$4     &  10$\pm$40     & -80$\pm$20      & -50$\pm$10 \\
LSR0519+4213  &   esdM3.5 &  55$\pm$27    &-200$\pm$40     &-340$\pm$130     & 110$\pm$70 \\
LSR0521+3425  &   M5.0 V  &  32$\pm$10    & -80$\pm$30     & -70$\pm$20      & -10$\pm$10 \\
LSR0522+3814  &   esdM3.0 &  45$\pm$22    &  30$\pm$40     &-360$\pm$170     &-130$\pm$60 \\
LSR0524+3358  &   sdM1.5  & 250$\pm$150   & 110$\pm$60     &-640$\pm$370     &  80$\pm$50 \\
LSR0527+3009  &   M5.0 V  &  40$\pm$13    & -10$\pm$30     & -30$\pm$10      &-120$\pm$40 \\
LSR0533+3837  &   sdM2.0  & 150$\pm$75    & 100$\pm$40     &-410$\pm$190     &  70$\pm$40 \\
LSR0539+4038  &   M8.0 Ve &  10$\pm$3     & -30$\pm$30     & -50$\pm$20      &   0$\pm$10 \\
LSR0541+3959  &   DA8     &  35$\pm$17    & -30$\pm$30     & -90$\pm$40      &   0$\pm$10 \\
LSR0549+2329  &   DC8     &  40$\pm$20    &  10$\pm$10     &-250$\pm$120     &  70$\pm$40 \\
LSR0556+1144  &   M5.5 V  &  10$\pm$3     &-160$\pm$30     & -70$\pm$10      & -10$\pm$10 \\
LSR0609+2319  &   sdM5.0  &  45$\pm$22    &  60$\pm$30     &-220$\pm$100     &  70$\pm$40 \\
LSR0618+1614  &   sdM2.0  &  85$\pm$42    & -10$\pm$40     &-270$\pm$120     & -70$\pm$30 \\
LSR0621+3652  &   esdK7   &  80$\pm$40    &-280$\pm$30     &-290$\pm$150     & -90$\pm$60 \\
LSR0627+0616  &   esdM1.5 &  80$\pm$40    & 260$\pm$80     &-260$\pm$160     &-170$\pm$90 \\
LSR0628+0529  &   M7.0 V  &  16$\pm$5     & -90$\pm$30     & -90$\pm$20      &   0$\pm$10 \\
LSR0646+3212  &   M5.5 V  &  60$\pm$20    &-150$\pm$80     &-170$\pm$50      &  70$\pm$20 \\
LSR0702+2154  &   M5.5 V  &  16$\pm$5     & -10$\pm$30     & -60$\pm$20      &  10$\pm$10 \\  
LSR0705+0506  &   sdM3.5  &  90$\pm$45    &  10$\pm$70     &-240$\pm$100     & -20$\pm$10 \\
LSR0721+3714  &   M5.5 Ve &  26$\pm$8     & -90$\pm$30     & -30$\pm$10      & -50$\pm$20 \\
LSR0731+0729  &   M5.0 V  &  24$\pm$8     &-120$\pm$40     &-120$\pm$30      & -10$\pm$20 \\
LSR0803+1547  &   sdM0.0  & 275$\pm$125   & 420$\pm$140    &-540$\pm$270     & -40$\pm$20 \\
LSR1722+1004  &   M4.0 V  &  70$\pm$23    & 130$\pm$50     &-200$\pm$70      & -60$\pm$20 \\
LSR1755+1648  &   sdM3.5  &  28$\pm$14    &  10$\pm$30     &   0$\pm$30      &-140$\pm$60 \\
LSR1757+0015  &   sdM4.5  &  75$\pm$37    &  90$\pm$50     &-150$\pm$80      & -80$\pm$40 \\
LSR1758+1417  &   DA10    &  18$\pm$9     &  30$\pm$20     & -70$\pm$30      &  30$\pm$20 \\
LSR1802+0028  &   esdM1.5 & 180$\pm$90    &-220$\pm$110    &-640$\pm$200     & -70$\pm$20 \\
LSR1806+1141  &   M4.0 V  &  45$\pm$15    &  70$\pm$40     & -80$\pm$40      & -40$\pm$20 \\
LSR1808+1134  &   M5.0 V  &  18$\pm$6     &  20$\pm$30     & -40$\pm$30      &  10$\pm$10 \\
LSR1809-0219  &   M4.5 V  &  22$\pm$7     &  50$\pm$40     & -30$\pm$20      & -30$\pm$10 \\
LSR1809-0247  &   M5.0 V  &  26$\pm$8     & -30$\pm$40     &-150$\pm$40      &   0$\pm$10 \\
LSR1817+1328  &   DA10    &  15$\pm$7     &  20$\pm$40     & -90$\pm$40      & -20$\pm$10 \\
LSR1820-0031  &   sdM2.0  &  85$\pm$42    & 100$\pm$60     &-200$\pm$100     & -40$\pm$20 \\
LSR1833+2219  &   M5.0 V  &  26$\pm$8     &  60$\pm$20     & -10$\pm$30      &   0$\pm$10 \\
LSR1835+3259  &   M9.0 Ve &   6$\pm$2     &   0$\pm$20     & -20$\pm$30      & -20$\pm$10 \\
LSR1836+1040  &   esdM0.5 & 100$\pm$50    & 170$\pm$150    &-430$\pm$170     & -30$\pm$10 \\
LSR1841+2421  &   M6.0 V  &  22$\pm$7     &  70$\pm$30     & -20$\pm$30      & -20$\pm$10 \\
LSR1843+0507  &   M5.5 V  &  38$\pm$12    & -10$\pm$30     &-100$\pm$30      &  60$\pm$20 \\
LSR1844+0947  &  (sd)M2.0V& 250$\pm$125   & 280$\pm$190    &-500$\pm$220     & 160$\pm$90 \\
\nodata       &  \nodata  &(100$\pm$50)   & (70$\pm$80)    &(-240$\pm$90)    & (60$\pm$40)\\
LSR1851+2641  &   M6.0 Ve &  20$\pm$6     & -60$\pm$20     & -40$\pm$30      & -10$\pm$10 \\
LSR1859+0156  &   M4.5 V  &  36$\pm$12    & -60$\pm$30     &-100$\pm$30      &-110$\pm$30 \\
LSR1914+2825A &   sdM0.0  & 200$\pm$110   & 390$\pm$240    &-320$\pm$140     &  30$\pm$30 \\
LSR1914+2825B &   sdM1.5  & 160$\pm$100   & 280$\pm$210    &-290$\pm$120     &-130$\pm$60 \\
LSR1918+1728  &   esdM3.0 & 125$\pm$62    & 170$\pm$140    &-350$\pm$110     &-110$\pm$50 \\
LSR1919+1438  &   M5.0 V  &  22$\pm$7     & -40$\pm$30     & -10$\pm$30      & -60$\pm$20 \\
LSR1922+4605  &   sdM0.0  & 200$\pm$90    & 500$\pm$230    &-140$\pm$60      &   0$\pm$10 \\
LSR1927+6802  &   M6.5 V  &  22$\pm$7     &  40$\pm$20     & -20$\pm$40      &  10$\pm$20 \\
LSR1928-0200A &   M3.5 V  &  85$\pm$27    & 100$\pm$70     &-340$\pm$90      & -70$\pm$20 \\
LSR1928-0200B &   M5.5 V  &  65$\pm$22    &  70$\pm$60     &-270$\pm$70      & -50$\pm$20 \\
LSR1933-0138  &   M3.0 V  &  50$\pm$16    &-140$\pm$30     &-170$\pm$30      &-180$\pm$60 \\
LSR1943+0941  &   M5.5 V  &  26$\pm$8     &  40$\pm$30     & -50$\pm$30      &   0$\pm$10 \\
LSR1945+4650A &   DA9     &  30$\pm$15    &  70$\pm$40     & -20$\pm$10      &  20$\pm$10 \\
LSR1945+4650B &   DC10    &  60$\pm$30    & 150$\pm$80     & -40$\pm$10      &  50$\pm$30 \\
LSR1946+0937  &   DA9     &  40$\pm$20    & -40$\pm$40     &-170$\pm$40      & -50$\pm$30 \\
LSR1946+0942  &   M3.5 V  &  90$\pm$30    &  30$\pm$50     &-240$\pm$50      &-130$\pm$50 \\
LSR1956+4428  &   esdM0.5 &  75$\pm$37    & 280$\pm$150    &-200$\pm$40      & -10$\pm$10 \\
LSR2000+0404  &   M5.5 V  &  22$\pm$7     &   0$\pm$30     &   0$\pm$30      & -60$\pm$20 \\
LSR2000+3057  &   M6.0 Ve &  16$\pm$5     &-100$\pm$30     &  20$\pm$40      &  20$\pm$10 \\
LSR2005+0835  &   sdK5    & 275$\pm$100   & 520$\pm$210    &-520$\pm$180     &-170$\pm$60 \\
LSR2009+5659  &   sdM2.0  &  55$\pm$27    &-220$\pm$110    &-130$\pm$40      & -30$\pm$10 \\
LSR2010+3938  &  (sd)M1.5V& 225$\pm$100   & 390$\pm$220    &-300$\pm$60      &-320$\pm$150 \\
\nodata       & \nodata   &(100$\pm$50)   &(140$\pm$100)   &(-250$\pm$40)    &(-150$\pm$70)\\
LSR2013+0417  &   sdK7    & 175$\pm$87    & 270$\pm$220    &-580$\pm$200     &-100$\pm$80 \\
LSR2017+0623  &   M5.0 V  &  40$\pm$13    &  80$\pm$40     & -70$\pm$40      & -60$\pm$20 \\
LSR2036+5059  &   sdM7.5  &  18$\pm$9     & -50$\pm$20     &-130$\pm$30      &-100$\pm$40 \\
LSR2044+1339  &   M5.0 Ve &  16$\pm$5     & -50$\pm$20     &   0$\pm$30      &   0$\pm$10 \\
LSR2050+7740  &   DC11    &  45$\pm$22    &-110$\pm$50     & -50$\pm$20      &  10$\pm$10 \\
LSR2059+5517  &   DC11    &  25$\pm$12    & -70$\pm$30     & -10$\pm$10      &   0$\pm$10 \\
LSR2107+3600  &   M4.5 V  &  60$\pm$20    & 170$\pm$70     &-160$\pm$40      &  40$\pm$10 \\
LSR2115+3804  &   esdK7   & 100$\pm$50    &-250$\pm$110    &-110$\pm$40      &  90$\pm$40 \\
LSR2117+7345  &   M6.0 V  &  20$\pm$6     &-100$\pm$20     &  30$\pm$40      &  10$\pm$10 \\
LSR2122+3656  &   esdM5.0 &  45$\pm$22    &-180$\pm$80     & -90$\pm$40      & -50$\pm$30 \\
LSR2124+4003  &   M6.5 V  &   7$\pm$2     & -40$\pm$10     & -60$\pm$30      &   0$\pm$10 \\
LSR2132+4754  &   M4.0 V  &  45$\pm$15    &-125$\pm$40     & -70$\pm$40      &  20$\pm$10 \\
LSR2146+5147  &  sdM1.0   & 200$\pm$100   &-540$\pm$290    &-280$\pm$50      & +10$\pm$10 \\
LSR2158+6117  &   M6.0 V  &  14$\pm$4     & -40$\pm$20     & -80$\pm$40      & -40$\pm$10 \\
LSR2205+5353  &   sdM1.0  & 100$\pm$50    & 270$\pm$120    &-160$\pm$50      &  10$\pm$10 \\
LSR2205+5807  &   esdM1.0 &  70$\pm$35    & -80$\pm$50     &-150$\pm$40      &-160$\pm$70 \\
LSR2251+4706  &   M6.5 V  &  35$\pm$11    & -90$\pm$30     &-130$\pm$40      &   0$\pm$10 \\
LSR2311+5032  &   M4.5 V  &  30$\pm$9     & -90$\pm$30     & -60$\pm$30      & -20$\pm$10 \\
LSR2311+5103  &   M7.5 V  &  26$\pm$8     &  50$\pm$20     &  10$\pm$40      &   0$\pm$10 \\
LSR2321+4704  &   esdM2.0 &  85$\pm$42    &-220$\pm$140    &-300$\pm$50      &  40$\pm$10 
\enddata
\tablenotetext{1}{Spectral class and subtype based on
spectroscopy. Parenthesis indicate possible change in spectral class
that would lower the star's components of velocity to more consistent 
values. The modified values are given in parenthesis one line below.}
\tablenotetext{2}{Spectroscopic distance estimate.}
\tablenotetext{3}{UVW components of the velocity relative to the local
standard of rest, calculated from the radial velocities, proper motions, 
and estimated distances.}
\end{deluxetable} 

\clearpage

\begin{deluxetable}{lrr}
\tabletypesize{\scriptsize}
\tablecolumns{3} 
\tablewidth{190pt} 
\tablecaption{Blackbody temperature fits for the white dwarfs}
\tablehead{
\colhead{Star} & 
\colhead{T$_{blackbody}$} &
\colhead{Spectral type}
}
\startdata 
LSR0310+6634  & 5000 & DC10 \\
LSR0401+5131  & 6500 & DC8  \\
LSR0541+3959  & 6000 & DA8  \\
LSR0549+2329  & 6000 & DC8  \\
LSR1758+1417  & 5250 & DA10 \\
LSR1817+1328  & 5000 & DA10 \\
LSR1945+4650A & 5750 & DA9  \\
LSR1945+4650B & 5000 & DC10 \\
LSR1946+0937  & 5750 & DA9  \\
LSR2050+7740  & 4750 & DC11 \\
LSR2059+5517  & 4750 & DC11 
\enddata     
\end{deluxetable} 


\clearpage

\begin{figure}
\plotone{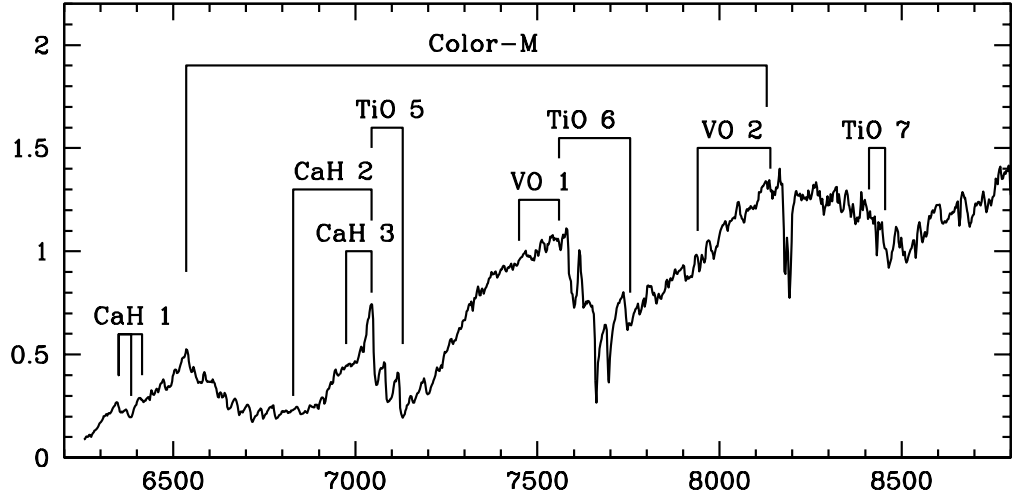}
\caption{\label{fig1} Illustration of the different spectral indices
cited in this paper, as defined in Table 1, here shown with the
spectrum of the star LSR1809-0247 (M5.0 V). The indices measure the
strengths of all the most prominent molecular features in the
6000\AA-9000\AA\ range.}
\end{figure}

\begin{figure}
\plotone{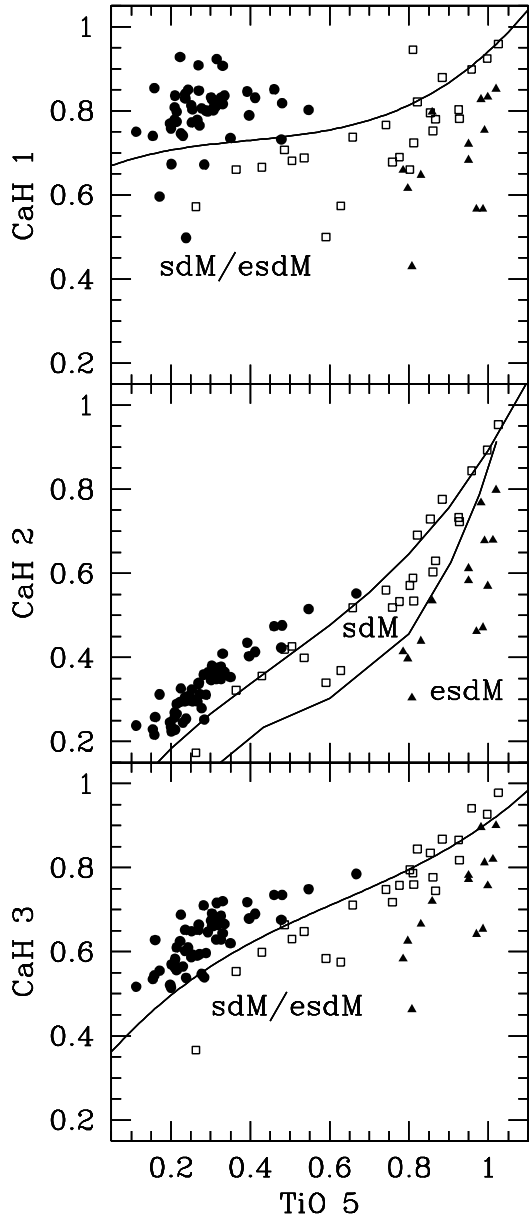}
\caption{\label{fig2} Separation of the high proper motion red dwarfs
into three metallicity classes (dwarfs dM, subdwarfs sdM, extreme subdwarfs 
esdM) based on the ratio of the CaH spectral indices with the TiO5 spectral
index. This classification scheme is based on the criteria defined by
G97. We use the CaH2/TiO5 ratio as the main criterion. The
three classes show up well separated by the CaH3/TiO5 ratio. The
CaH1/TiO5 ratio, on the other hand, does not discriminate as well
between the dM and sdM classes; the shorter dynamic range of the CaH1
index makes classification with the CaH1/TiO5 ratio more sensitive to
measurement errors.}
\end{figure}

\begin{figure}
\plotone{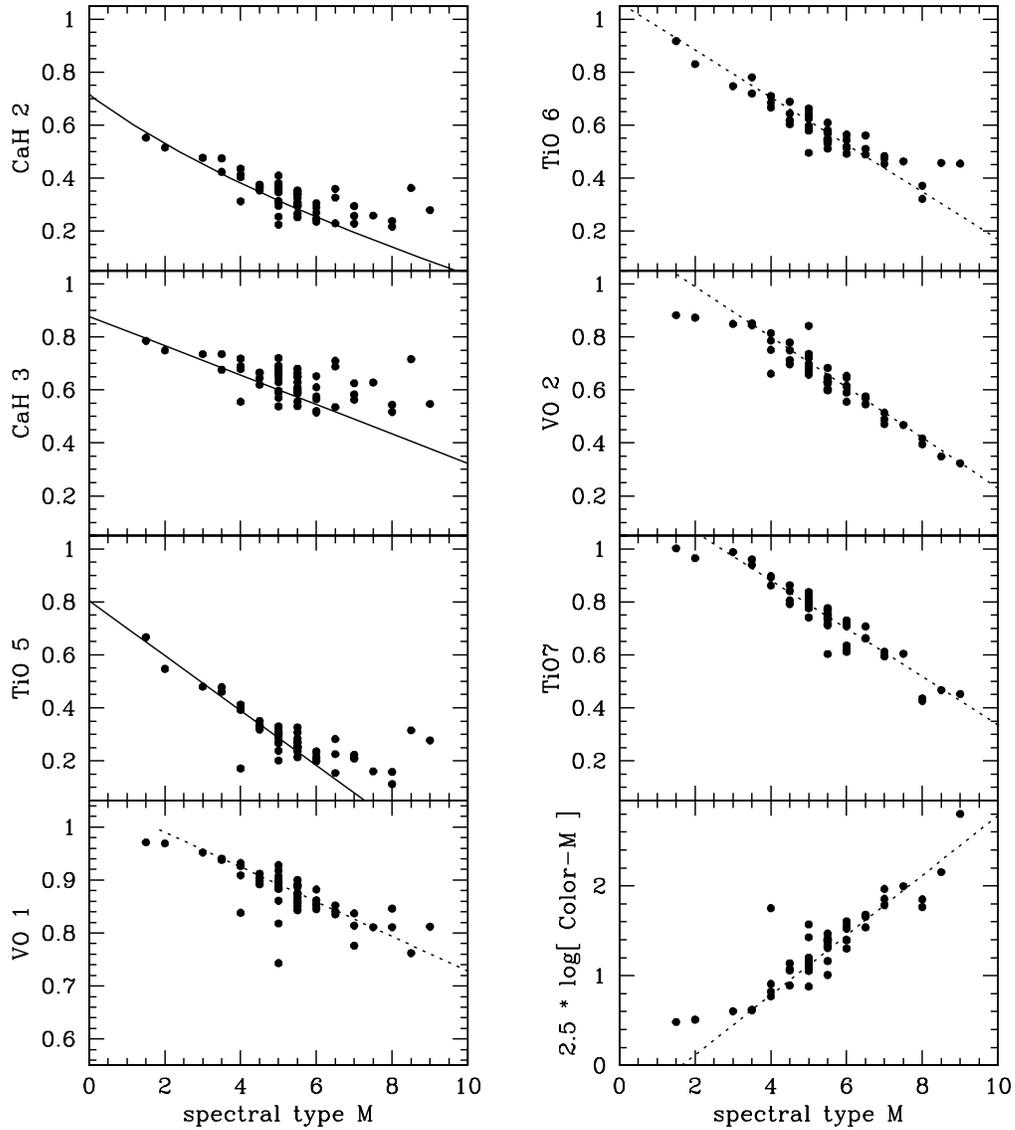}
\caption{\label{fig3}
Correlation between spectral type and the different spectral indices,
for the 54 M dwarf stars classified in this paper. The relationships
determined by G97 for the CaH2, CaH3, and TiO5 indices are
plotted in continuous lines. Our new relationships for the other indices
are plotted as dotted lines. Classification of early-type M dwarfs is
more reliable when the blueward indices (CaH2, TiO5) are used, while
late-type M dwarfs are better classified with our newly defined
redward indices (VO 2, TiO7). Negative values correspond to K dwarf
spectra, with -2$\rightarrow$K5V and -1$\rightarrow$K7V.}
\end{figure}

\begin{figure}
\plotone{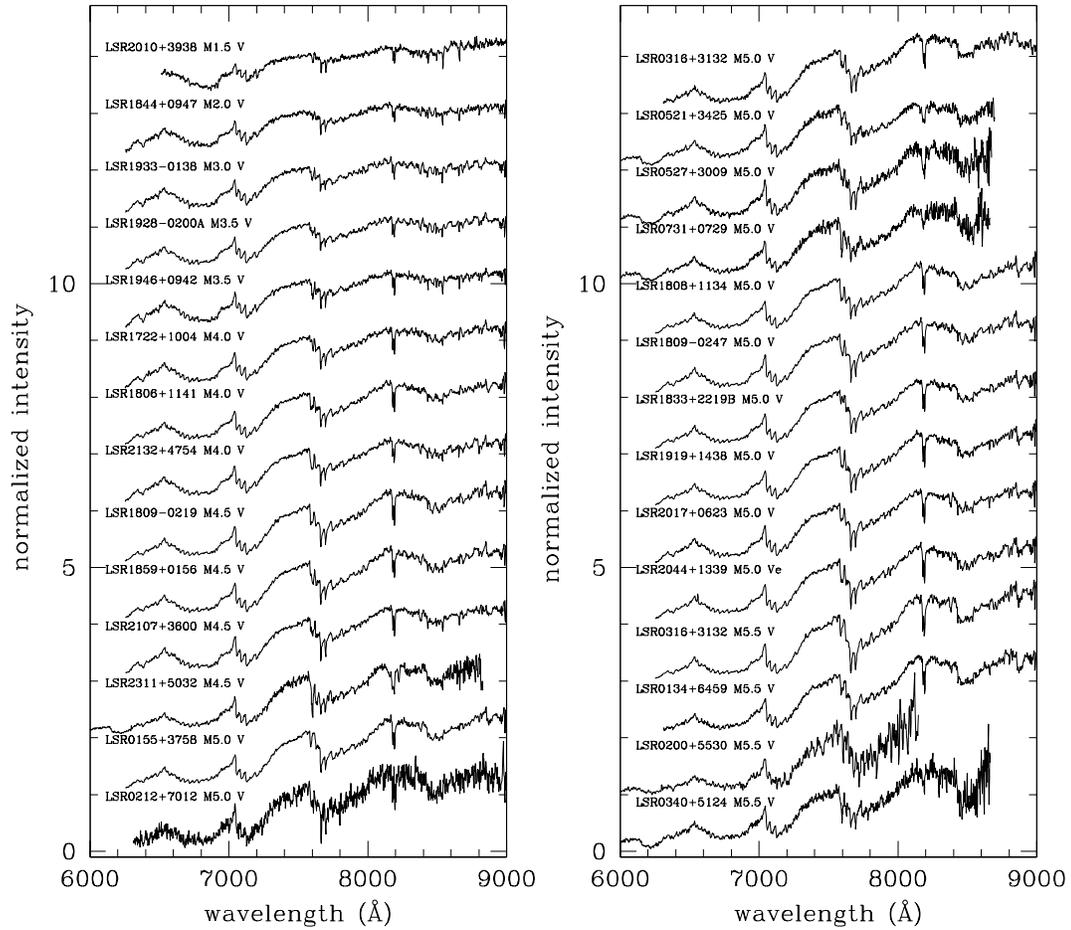}
\caption{\label{fig4} Spectral sequence for M dwarf stars classified
in this paper. Spectra have been normalized to 1.0 at 7500\AA\ and
shifted vertically by integer values for clarity. The later-type stars
are displayed on Figure 5.}
\end{figure}

\begin{figure}
\plotone{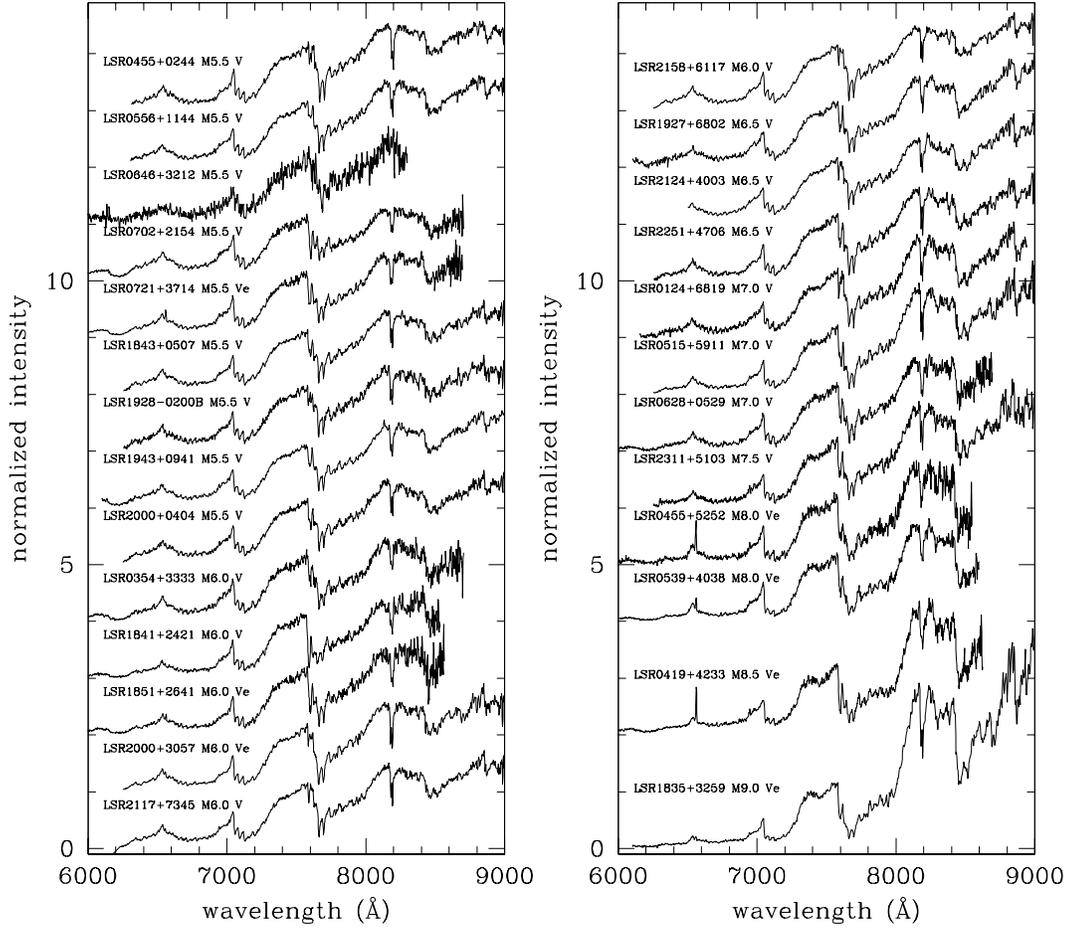}
\caption{\label{fig5}Spectral sequence for M dwarf stars classified in
this paper, continued from Figure 4.}
\end{figure}

\begin{figure}
\plotone{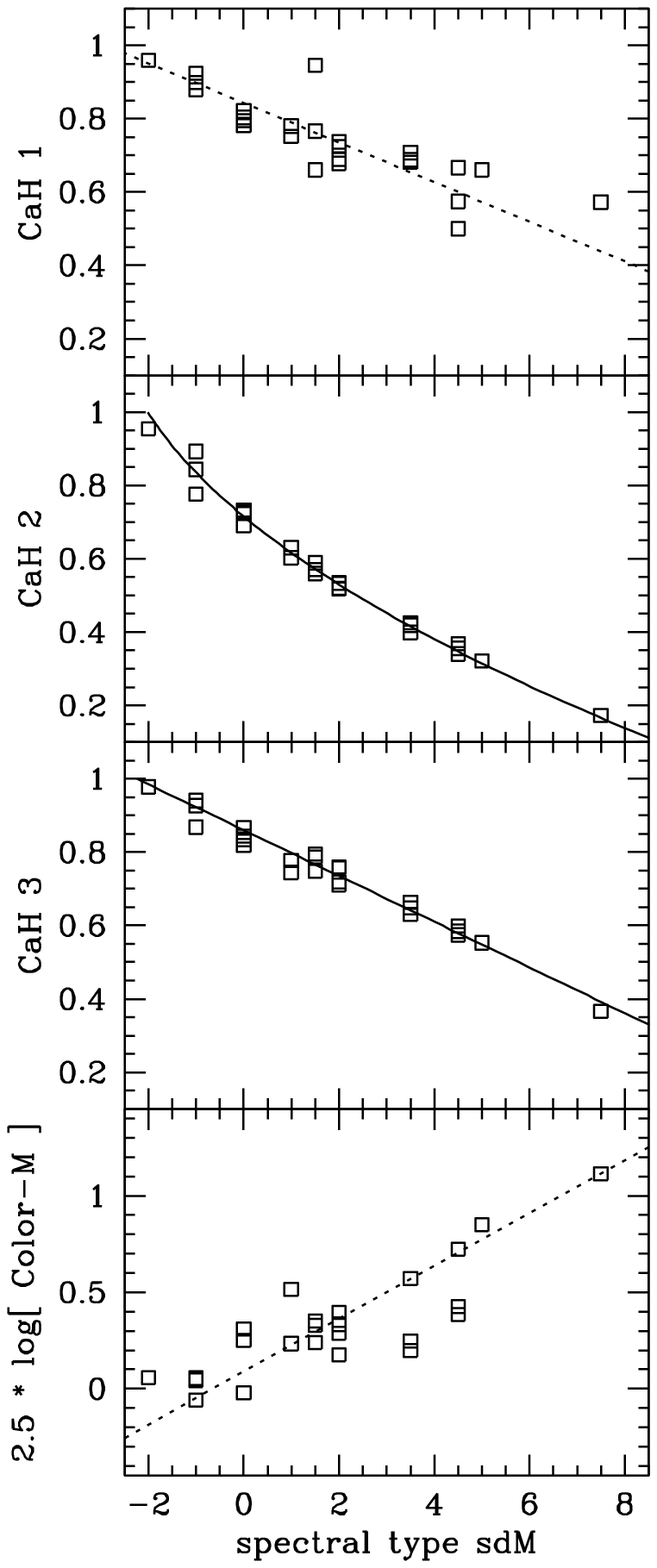}
\caption{\label{fig6}
Classification of the subdwarfs based on spectral
features. Classification is based of the CaH2 and CaH3 spectral
indices. The relationships determined by G97 are plotted in full 
lines. Values of the CaH1 index and of the Color-M index are plotted
against spectral type for comparison. Negative values correspond to sdK
dwarf spectra, with -2$\rightarrow$sdK5 and -1$\rightarrow$sdK7.}
\end{figure}

\begin{figure}
\plotone{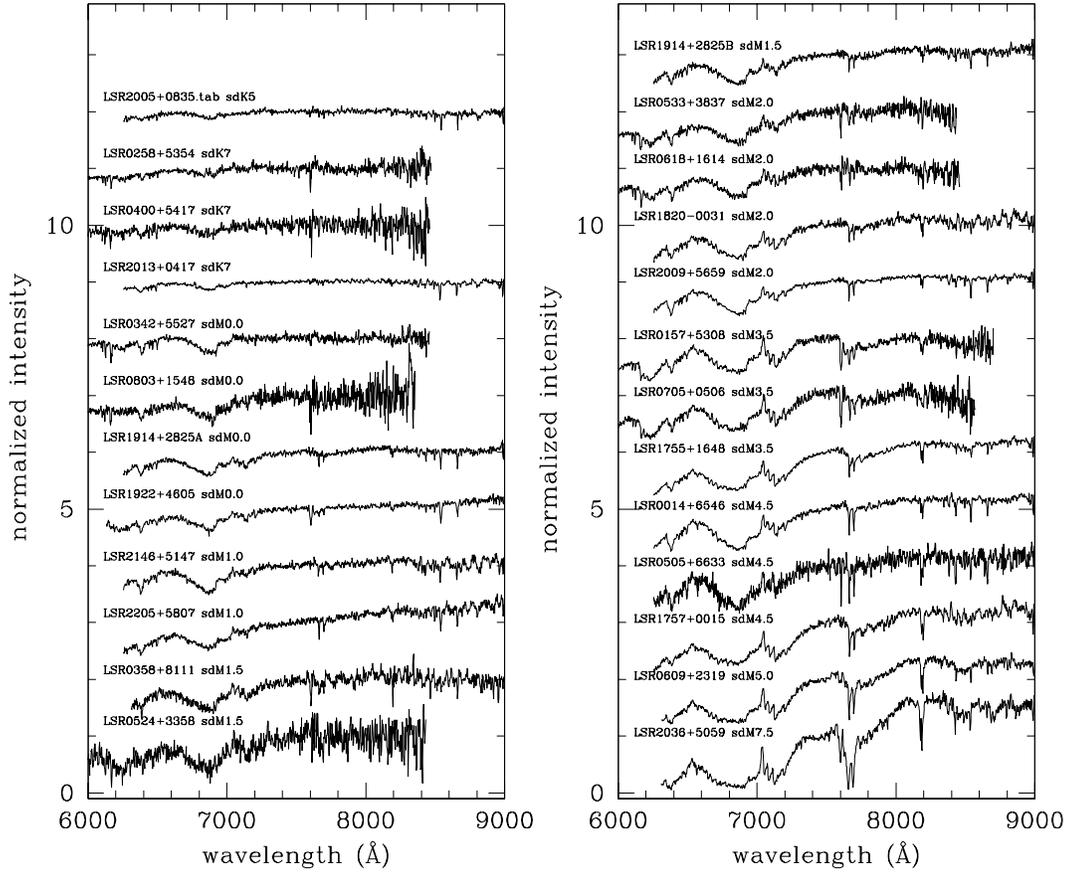}
\caption{\label{fig7} Spectral sequence for subdwarf (sdM) stars
classified in this paper. Spectra have been normalized to 1.0 at
7500\AA\ and shifted vertically by integer values for clarity.}
\end{figure}

\begin{figure}
\plotone{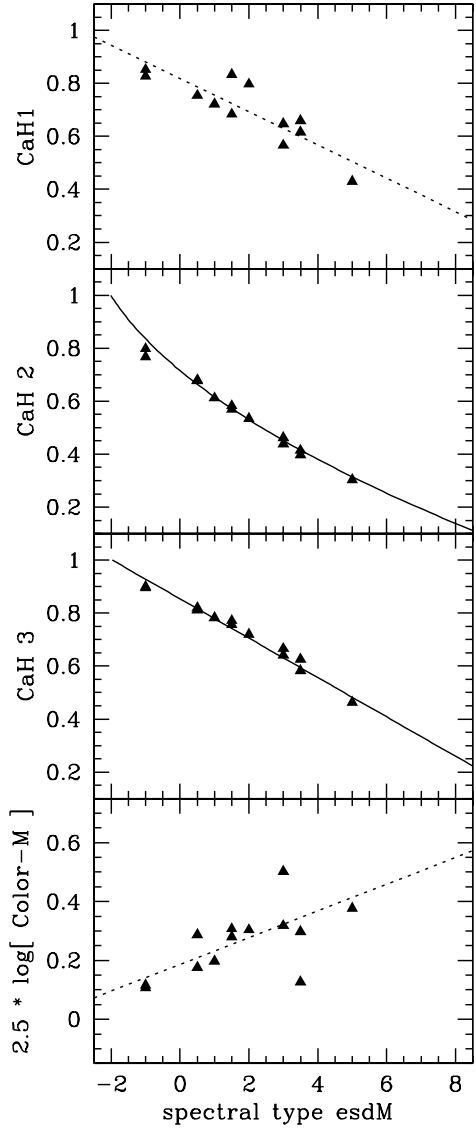}
\caption{\label{fig8} Classification of the extreme subdwarfs
based on spectral features. Classification is based of the CaH2 and
CaH3 spectral indices. The relationships determined by G97 are
plotted in full lines. Values of the CaH1 index and of the slope of
the continuum index are plotted against spectral type for
comparison. Negative values correspond to esdK dwarf spectra, with
-1$\rightarrow$esdK7.}
\end{figure}

\begin{figure}
\plotone{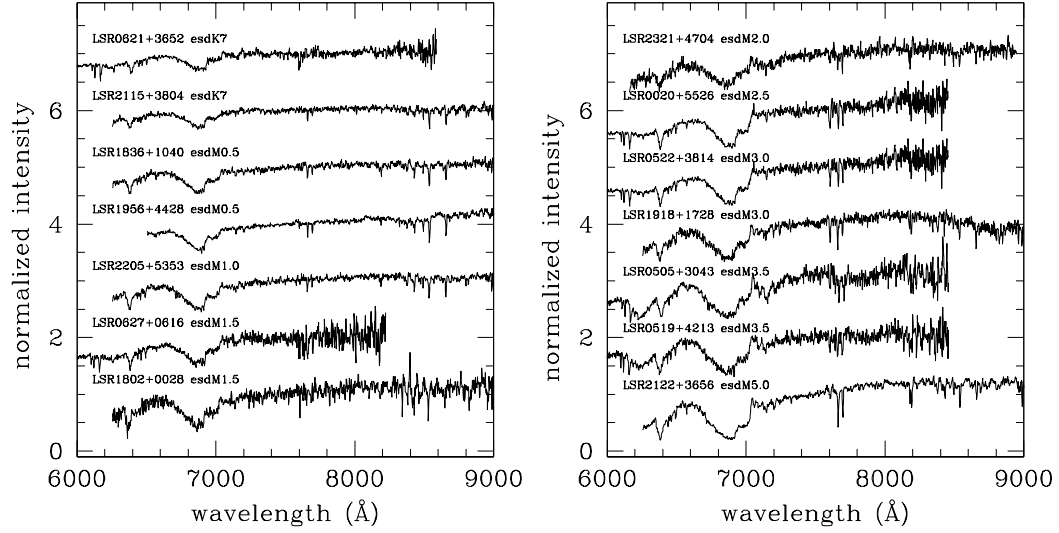}
\caption{\label{fig9} Spectral sequence for extreme subdwarf (esdM)
stars classified in this paper. Spectra have been normalized to 1.0 at
7500\AA\ and shifted vertically by integer values for clarity. Several
spectra are very noisy because we used a spectroscopic setup optimized
for the observation of red dwarfs.}
\end{figure}

\begin{figure}
\plotone{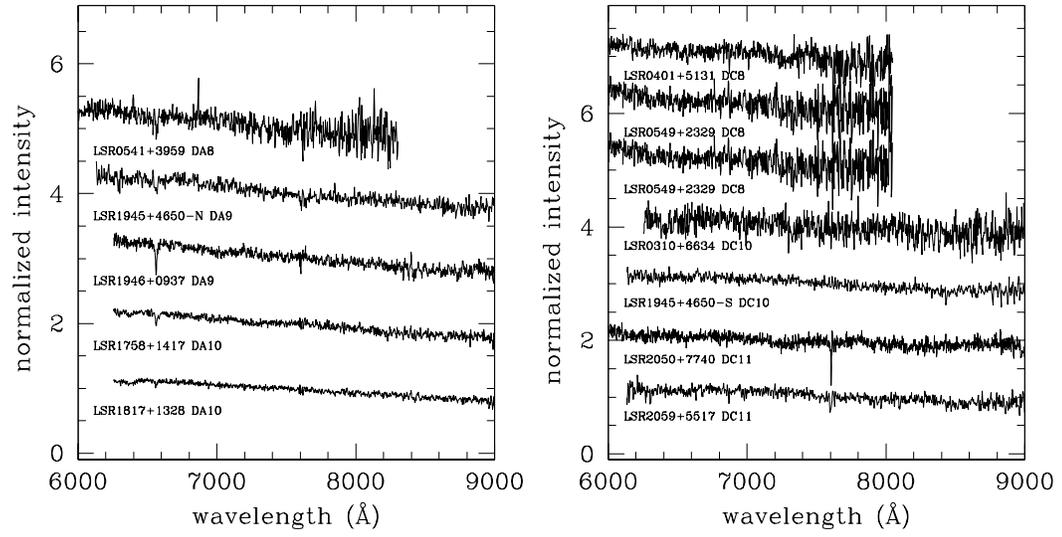}
\caption{\label{fig10} Spectra of the white dwarfs found in the sample of
new high proper motion stars. All the spectra are normalized to 1.0 at
7500\AA, and shifted by integer values on the vertical scale for
clarity.}
\end{figure}

\begin{figure}
\plotone{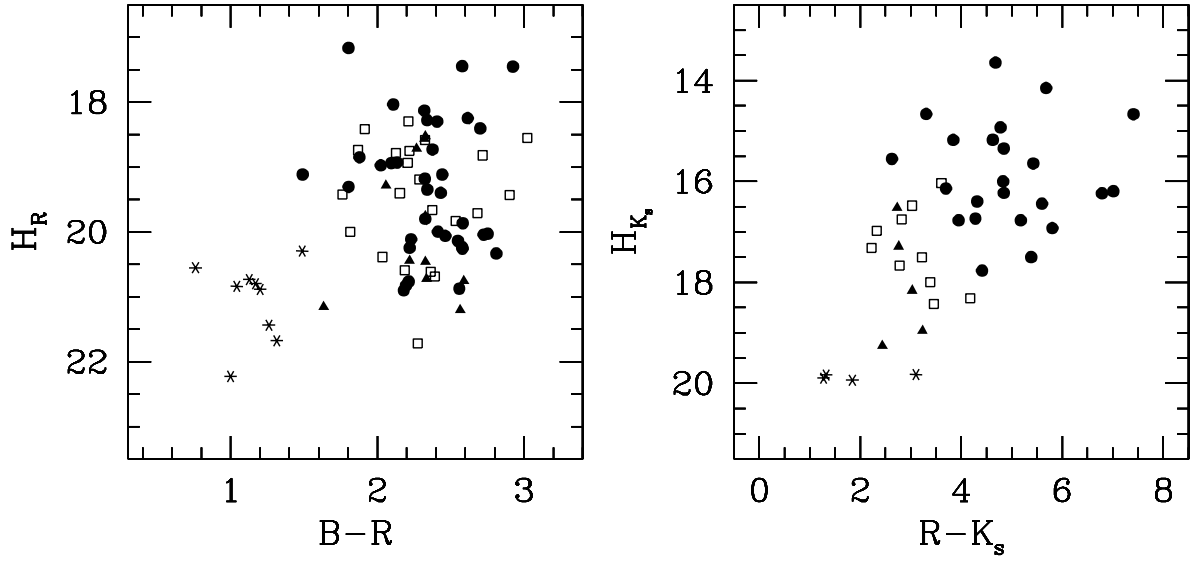}
\caption{\label{fig11} Reduced proper motion diagrams for the stars
classified in this paper. The reduced proper motion terms are derived
from H$_{\rm R}$ = R + $\log{\mu}$ + 5, and H$_{\rm K_s}$ = ${\rm
K_s}$ + $\log{\mu}$, where $\mu$ is the proper motion of the star. Red
dwarfs are represented by filled circles, subdwarfs (sdM) by open
squares, extreme subdwarfs (esdM) by filled triangles, and white dwarf
by asterisks.}
\end{figure}

\begin{figure}
\plotone{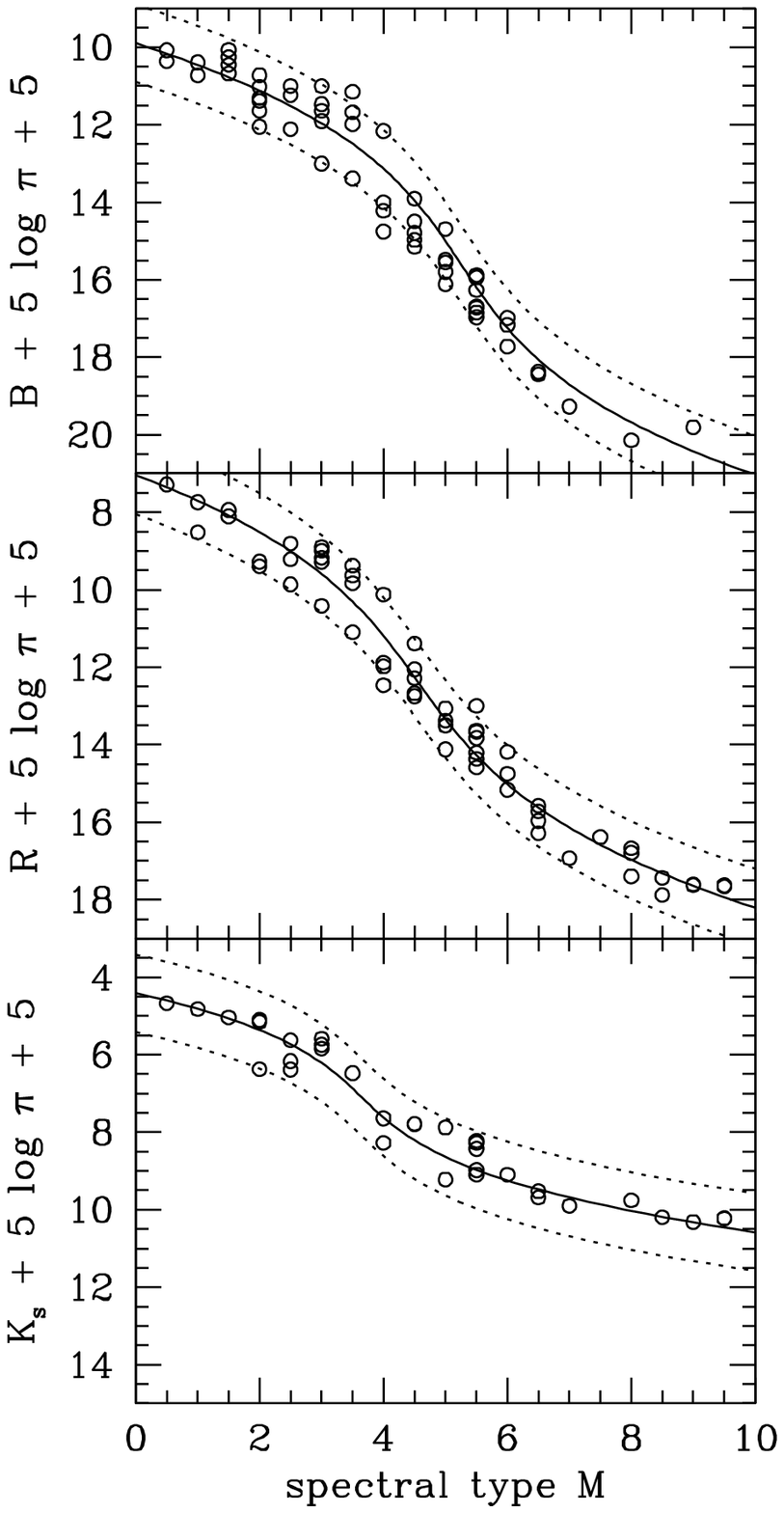}
\caption{\label{fig12} Empirical spectral-type / absolute magnitude
relationship for a sample of nearby M dwarfs with assigned spectral
types and measured astrometric parallaxes ($\pi$). The observed B and R
magnitudes are those recorded on the IIIaJ (blue) and IIIaF (red)
POSS-II plates as listed in the Second Guide Star Catalog (GSC2.2.1),
the ${\rm K_s}$ magnitudes are from the 2MASS Second incremental
Release. The relationships (continuous lines) are determined from
third order polynomial fits. Most stars fall within 1 magnitude of the
adopted relationship (dashed lines).}
\end{figure}

\begin{figure}
\plotone{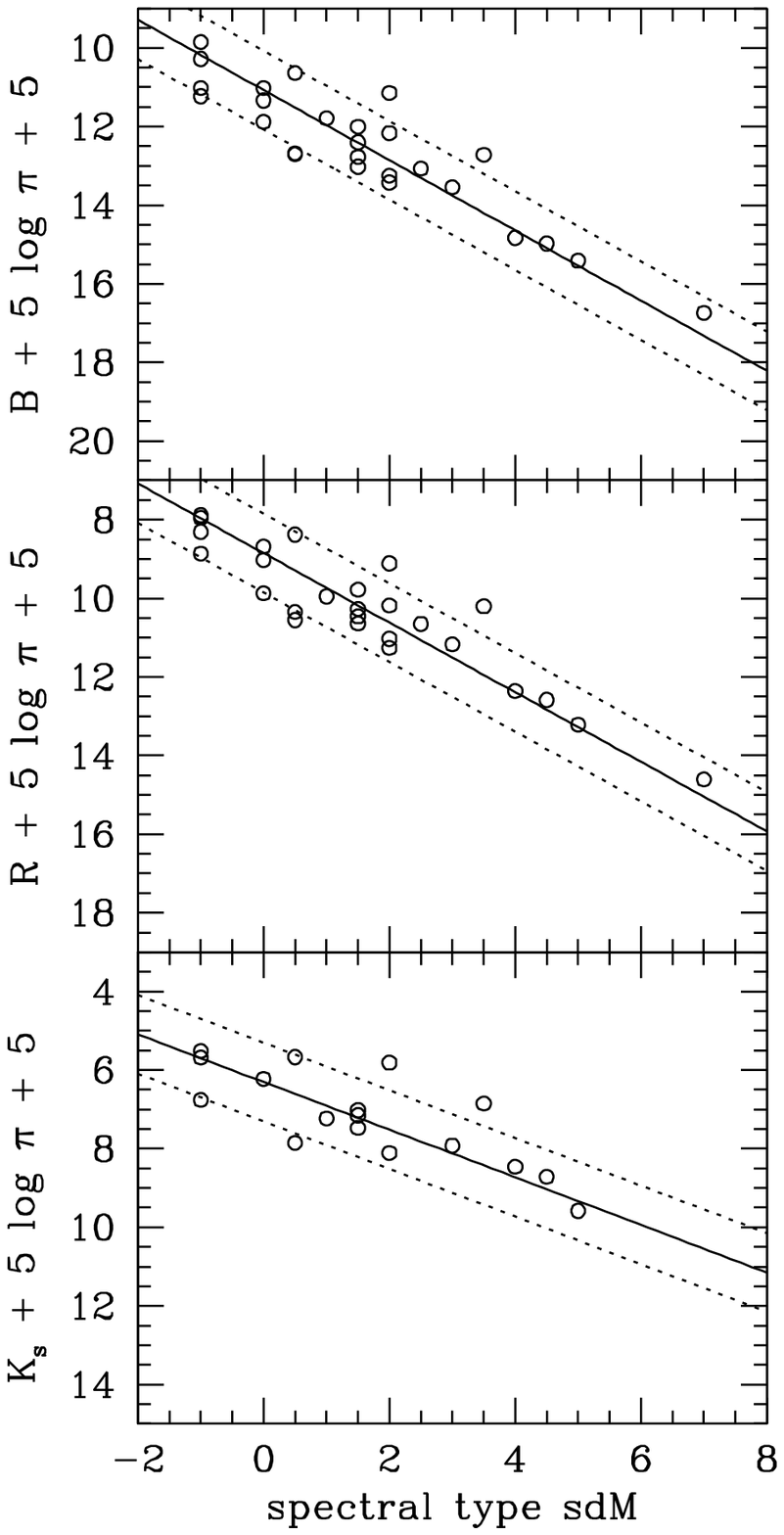}
\caption{\label{fig13} Empirical spectral-type / absolute magnitude
relationship for a sample of sdM dwarfs with assigned spectral types
and measured astrometric parallaxes ($\pi$). The observed B and R
magnitudes are those recorded on the IIIaJ (blue) and IIIaF (red)
POSS-II plates as listed in the Second Guide Star Catalog (GSC2.2.1),
the ${\rm K_s}$ magnitudes are from the 2MASS Second incremental
Release. The linear relationships (continuous lines) are determined
from first order polynomial fits. Dashed lines mark the 1 magnitude
scatter.}
\end{figure}

\begin{figure}
\plotone{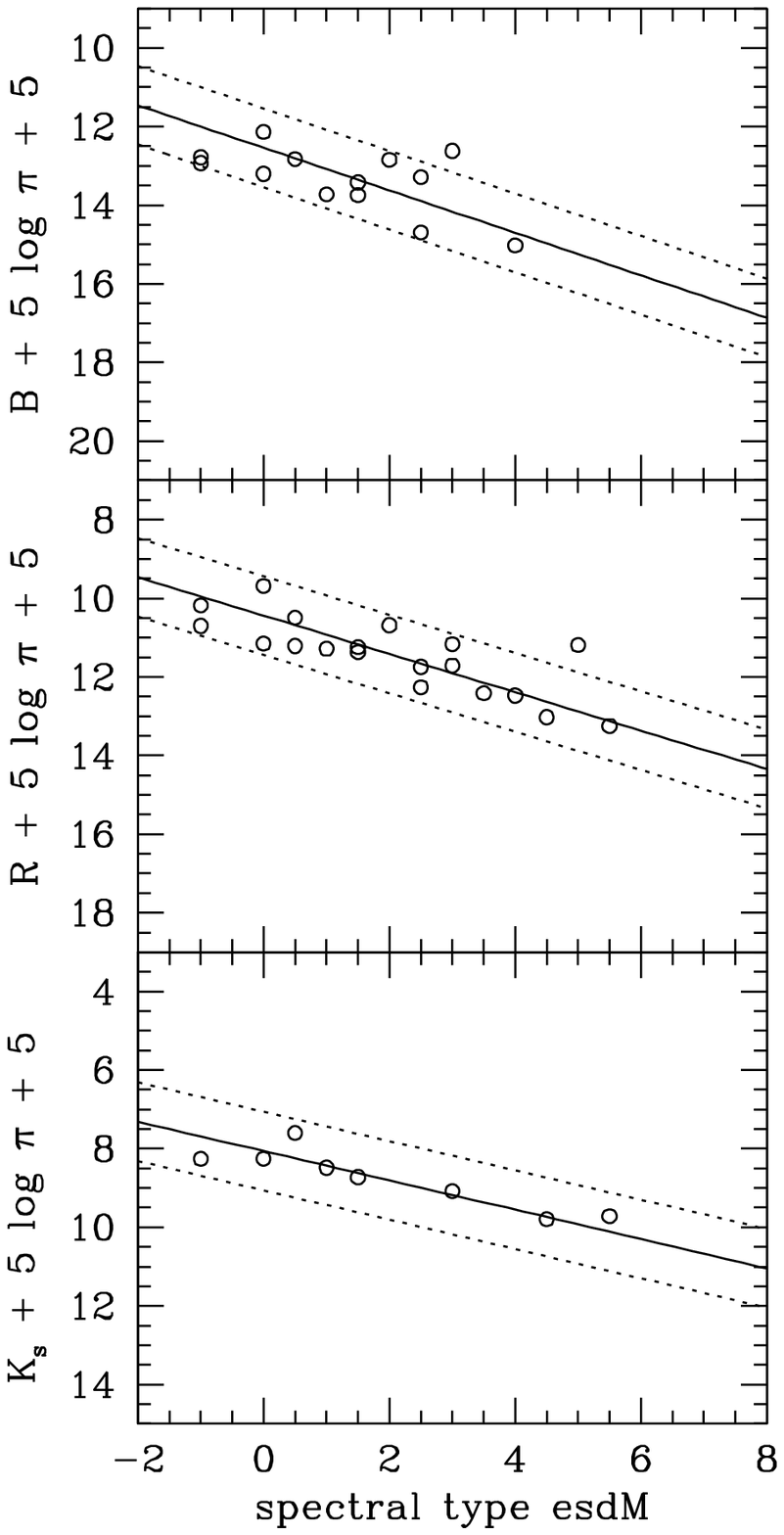}
\caption{\label{fig14} Empirical spectral-type / absolute magnitude
relationship for a sample of esdM dwarfs with assigned spectral types
and measured astrometric parallaxes ($\pi$). The observed B and R
magnitudes are those recorded on the IIIaJ (blue) and IIIaF (red)
POSS-II plates as listed in the Second Guide Star Catalog (GSC2.2.1),
the ${\rm K_s}$ magnitudes are from the 2MASS Second incremental
Release. The linear relationships are determined from first order
polynomial fits. Dashed lines mark the 1 magnitude scatter.}
\end{figure}

\begin{figure}
\plotone{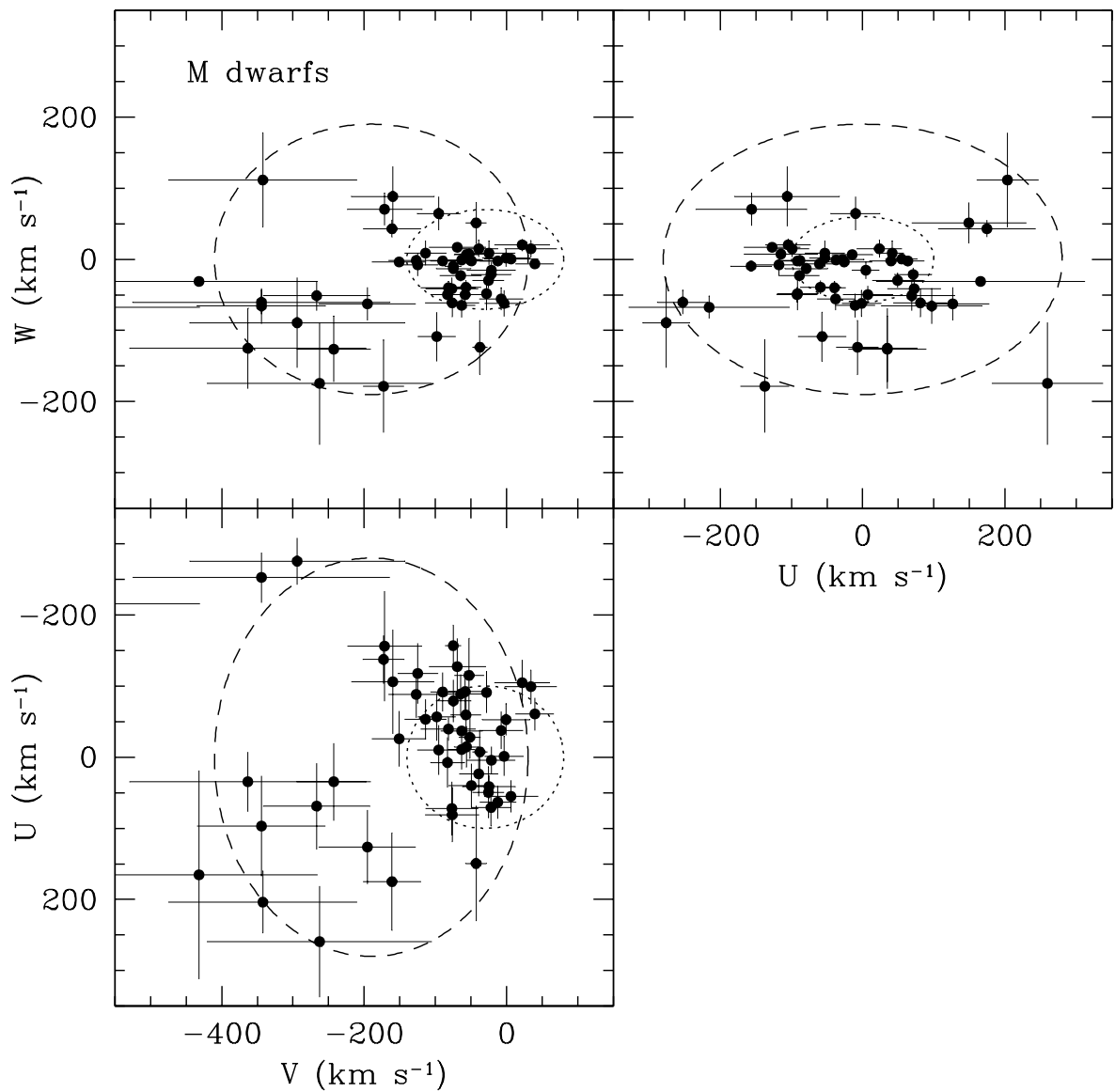}
\caption{\label{fig15} UVW velocity space distribution of the newly
classified M dwarfs. The dotted line shows the $2\sigma$ dispersion of
local  disk stars and the dashed line the $2\sigma$ dispersion of
local halo stars, based on observations compiled by \citet{CB00}. A
significant number of M dwarfs have kinematics that are more
consistent with the halo than with the disk.}
\end{figure}

\begin{figure}
\plotone{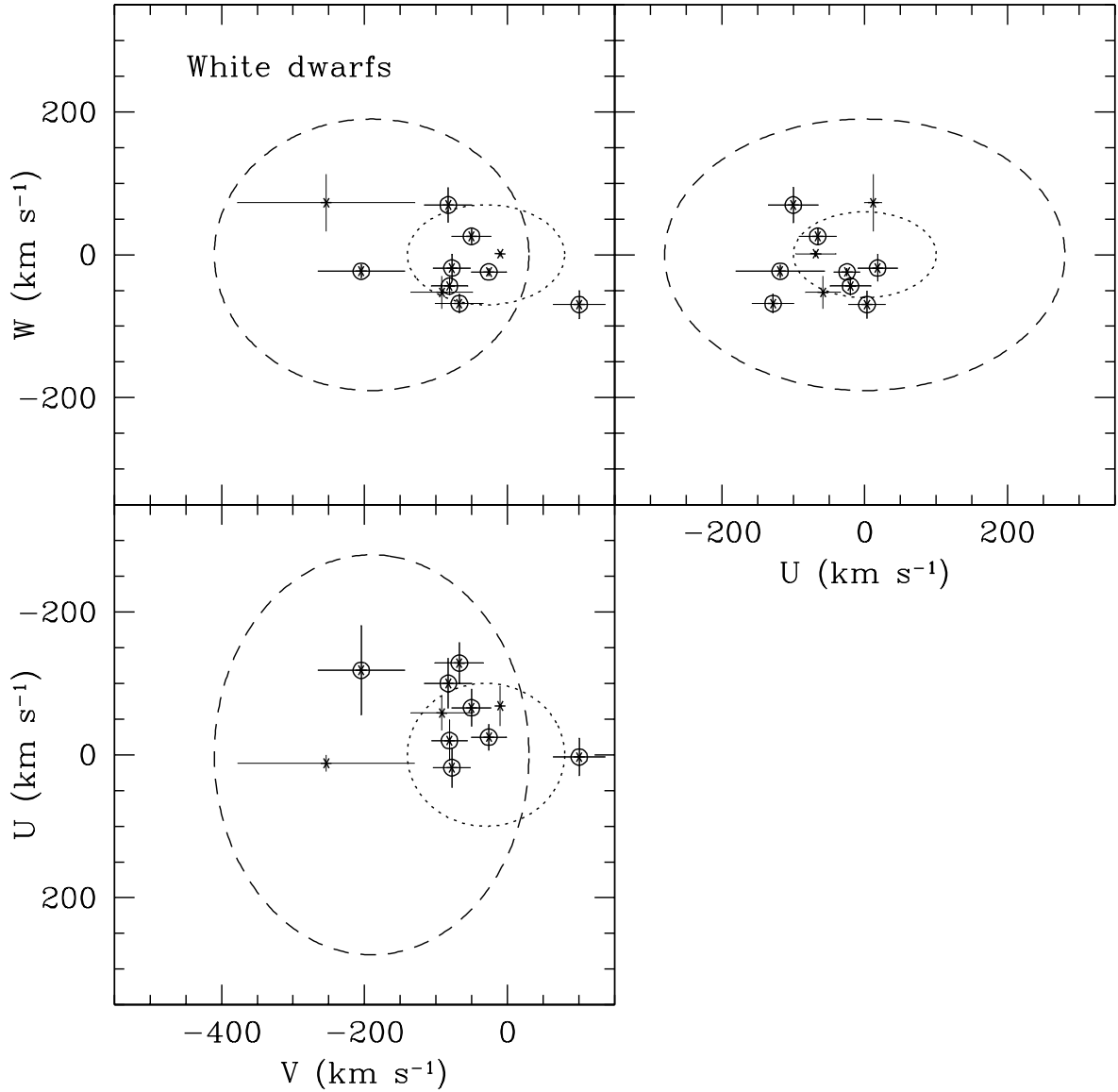}
\caption{\label{fig16} Same as in Figure 15, but for the new white
dwarfs identified in our sample. Circled asterisks indicate those
stars for which we do not have radial velocities; heliocentric radial
velocity HRV=0 is then assumed. The distribution appears to be very
similar to that of the M dwarfs (see Fig.15) with a few white dwarfs
falling in the region more typical for local halo stars.}
\end{figure}

\begin{figure}
\plotone{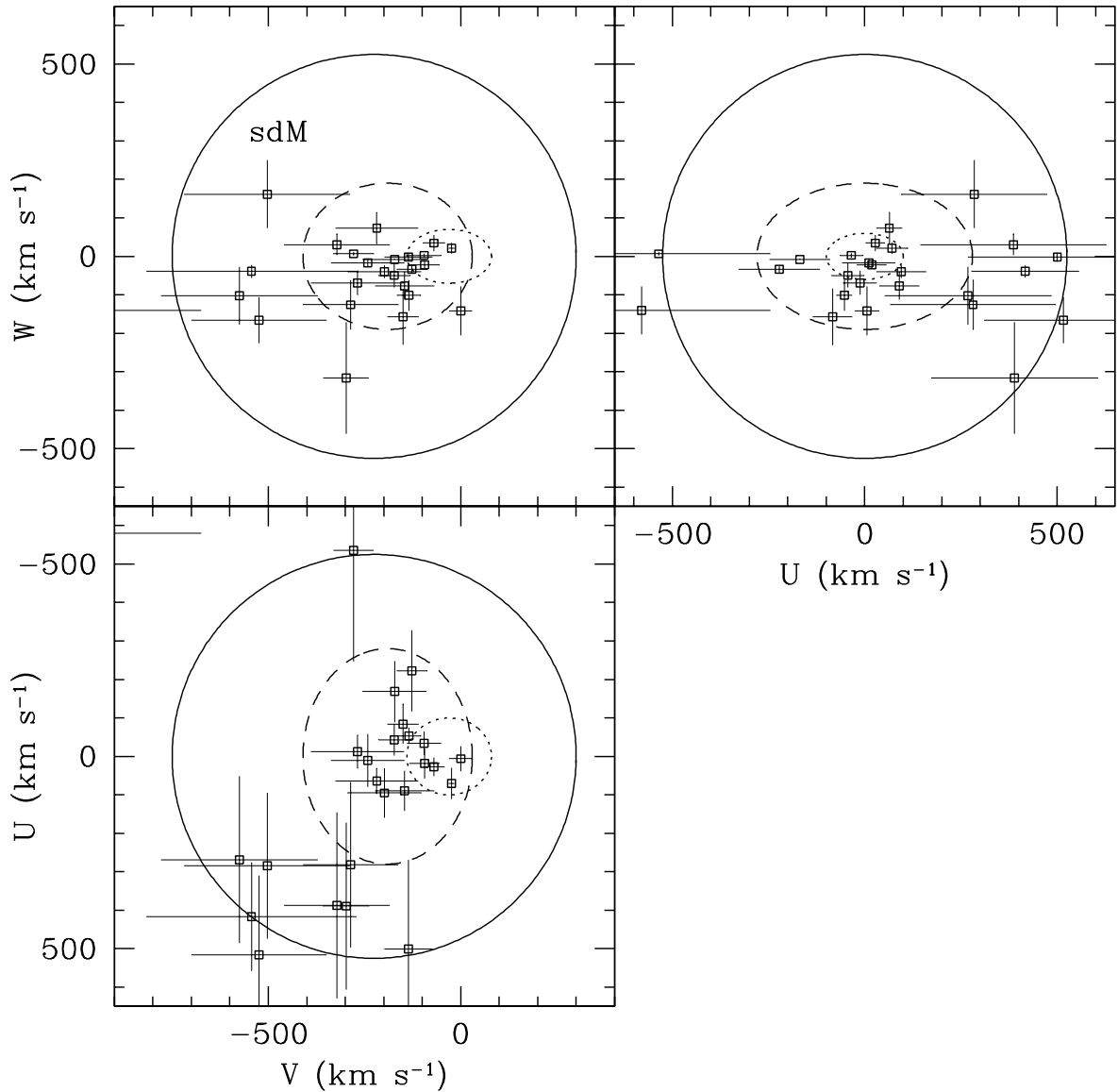}
\caption{\label{fig17} UVW velocity space distribution of the newly
classified subdwarfs. The dotted line shows the
$2\sigma$ dispersion of local  disk stars and the dashed line the
$2\sigma$ dispersion of local halo stars, based on observations
compiled by \citet{CB00}. The full line shows the limits within which
stars are gravitationally bound to the Galaxy, as calculated from the
Galactic model of \citet{DC95}. The distribution is significantly
different from that of the M dwarfs and white dwarfs (see Figs.15-16),
and is clearly consistent with a kinematics similar to halo stars.}
\end{figure}
\begin{figure}

\plotone{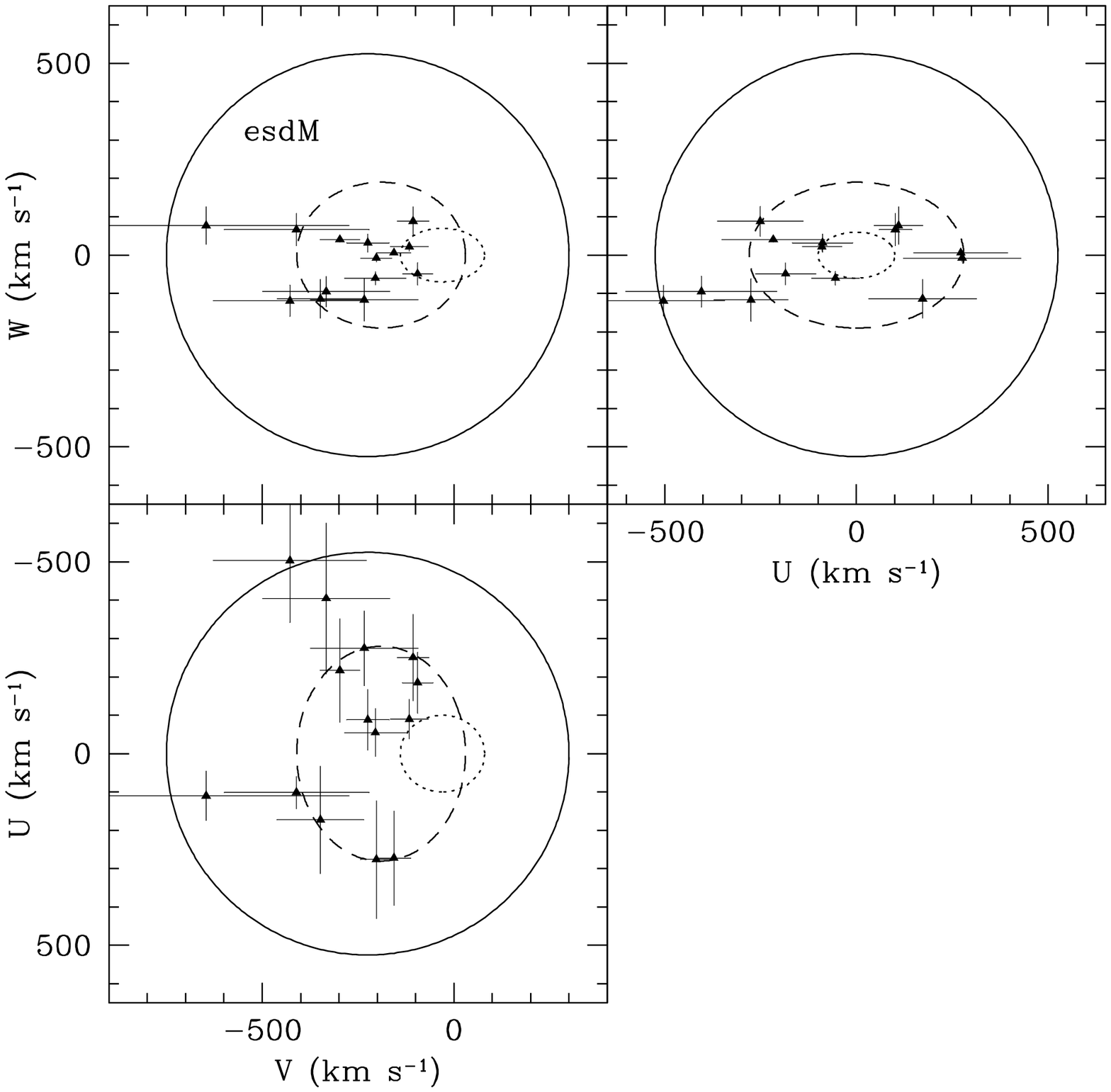}
\caption{\label{fig18} Same as in Figure 17, but for the extreme
subdwarfs in our sample. Again, the distribution of esdK/esdM stars is
significantly different from that of the M dwarfs and white dwarfs
(see Figs.15-16), and is clearly consistent with a kinematics similar
to halo stars. The velocity space distribution of extreme subdwarfs
does not, however, appear to be different from that of the subdwarfs
(see Fig.17).}
\end{figure}


\begin{thebibliography}{}
\bibitem[Chiba \& Beers(2000)]{CB00} 
   Chiba, M., \& Beers, T. 2000, \aj, 119, 2843
\bibitem[Cruz \& Reid(2002)]{CR02} 
   Cruz, K. L., \& Reid, I. N. 2002, \aj, 123, 2828
\bibitem[Dahn {\it et al.}(2002)]{D02}
   Dahn, C. C., Harris, H. C., Vrba, F. J., Guetter, H. H., Canzian, B.,
   Henden, A. A., Levine, S. E., Luginbuhl, C. B., Monet, A. K. B.,
   Monet, D. G., Pier, J. R., Stone, R. C., Walker, R. L., Burgasser,
   Adam J., Gizis, J. E., Kirkpatrick, J. D., Liebert, J., \& Reid,
   I. N. 2002, \aj, 124, 1170
\bibitem[Dauphole \& Colin(1995)]{DC95}
   Dauphole, B., \& Colin, J. 1995, \aap, 300, 117
\bibitem[Dehnen \& Binney(1998)]{DB98}
   Dehnen, W., \& Binney, J. J. 1998, \mnras, 298, 387
\bibitem[Delfosse {\it et al.}(1999)]{DTFEBFKT99}
   Delfosse, X., Tinney, C. G., Forveille, T., Epchtein, N.,
   Borsenberger, J., Fouqu\'e, P., Kimeswenger, S., \& Tiph\`ene,
   D. 1999, \aaps, 135, 41
\bibitem[Delfosse {\it et al.}(2001)]{DFMGBCAEFST01}
   Delfosse, X., Forveille, T., Mart\'{\i}n, E. L., Guibert, J.,
   Borsenberger, J., Crifo, F., Alard, C., Epchtein, N., Fouqu\'e, P.,
   Simon, G., \& Tajahmady, F. 2001, \aap, 366, L13
\bibitem[ESA(1997)]{E97}
   ESA, 1997, The Hipparcos and Tycho Catalogues, ESA SP-1200
   ({\it CDS-ViZier catalog number I/239})
\bibitem[Gizis(1997)]{G97}
   Gizis, J. E. 1997, \aj, 113, 806 (G97)
\bibitem[Gizis \& Reid(1997)]{GR97}
   Gizis, J. E., \& Reid, I. N. 1997, \pasp, 109, 849
\bibitem[Gizis {\it et al.}(2000)]{GMRKLW00}
   Gizis, J. E., Monet, D. G., Reid, I. N., Kirkpatrick, J. D.,
   Liebert, J., \& Williams, R. J. 2000, \aj, 120, 1085
\bibitem[Henry {\it et al.}(1997)]{HWBG97}
   Henry, T. J., Ianna, P. A., Kirkpatrick, J. D., \& Jahreiss,
   H. 1997, \aj, 114, 388
\bibitem[Henry {\it et al.}(2002)]{HWBG02}
   Henry, T. J., Walkowicz, L. M., Barto, T. C., \& Golimowski,
   D. A. 2002, \aj, 123, 2002
\bibitem[Hawley {\it et al.}(2002)]{H02}
   Hawley, S. L., {\it et al.} 2002, \aj, 123, 3409
\bibitem[Hurley \& Shara(2002)]{HS02}
   Hurley, J. R., Shara, \& M. M. 2002, \apj, 570, 184
\bibitem[Jahreiss {\it et al.}(2001)]{JSML01}
   Jahreiss, H., Scholz, R., Meusinger, H., \& Lehmann, I. 2001, \aap,
   370, 967
\bibitem[Kirkpatrick, Henry, \& McCarthy(1991)]{KHM91}
   Kirkpatrick, J. D., Henry, T. J., \& McCarthy, D. W. 1991, \apjs,
   77, 417
\bibitem[Kirkpatrick, Henry, \& Simons(1995)]{KHS95}
   Kirkpatrick, J. D., Henry, T. J., Simons, D. A. 1995, \aj, 109, 797
\bibitem[Kirkpatrick {\it et al.}(1999)]{KRLCNBDMGS99}
   Kirkpatrick, J. D., Reid, I. N., Liebert, J., Cutri, R. M., Nelson,
   B., Beichman, C. A., Dahn, C. C., Monet, D. G., Gizis, J. E., \&
   Skrutskie, M. F. 1999, \apj, 519, 802
\bibitem[Kroupa(2002)]{K02}
   Kroupa, P. 2002, \mnras, 330, 707
\bibitem[Lepine, Shara, \& Rich(2002a)]{LSR02a}
   L\'epine, S., Shara, M. M., \& Rich, R. M. 2002a, \aj, 123, 3434
\bibitem[Lepine, Shara, \& Rich(2002b)]{LSR02b}
   L\'epine, S., Shara, M. M., \& Rich, R. M. 2002b, \aj, 124, 1190
\bibitem[Luyten(1925)]{L25}
   Luyten W. J. 1925, \apj, 62, 8
\bibitem[Luyten(1979)]{L79}
   Luyten W. J. 1979, LHS Catalogue: a catalogue of stars with proper
   motions exceeding 0.5" annually, University of Minnesota,
   Minneapolis ({\it CDS-ViZier catalog number I/87B})
\bibitem[Luyten(1980)]{L80}
   Luyten W. J. 1980, New Luyten Catalogue of stars with proper motions
   larger than two tenths of an arcsecond (NLTT), University of
   Minnesota, Minneapolis ({\it CDS-ViZier catalog number I/98A})
\bibitem[Mart\'{\i}n {\it et al.}(1999)]{MDBGFZ99}
   Mart\'{\i}n, E. L., Delfosse, X., Basri, G., Goldman, B.,
   Forveille, T., Zapatero Osorio, M. R. 1999, \aj, 118, 2466
\bibitem[Massey {\it et al.}(1988)]{MSBA88}
   Massey, P., Strobel, K., Barnes, J. V., Anderson, E. 1988, \apj,
   328, 315
\bibitem[Massey \& Gronwall(1990)]{MG90}
   Massey, P., \& Gronwall, C. 1990, \apj, 358, 344
\bibitem[Oppenheimer {\it et al.}(2001)]{OHDHS01}
   Oppenheimer, B. R., Hambly, N. C., Digby, A. P., Hodgkin, S. T., \&
   Saumon, D. 2001, Science, Volume 292, Issue 5517, pp. 698-702
\bibitem[Phan-Bao {\it et al.}(2001)]{PGCDFBEFS01}
   Phan-Bao, N., Guibert, J., Crifo, F., Delfosse, X., Forveille, T.,
   Borsenberger, J., Epchtein, N., Fouqu\'e, P., \& Simon, G. 2001,
   \aap, 380, 590
\bibitem[Raboud {\it et al.}(1998)]{RGMFU98}
   Raboud, D., Grenon, M., Martinet, L., Fux, R., \& Udry, S. 1998
   \aap, 335, L61
\bibitem[Reid, Hawley, \& Gizis(1995)]{RHG95}
   Reid, I. N., Hawley, S. L., \& Gizis, J. E. 1995, \aj, 110, 1838
\bibitem[Reid, Liebert, \& Schmidt(2001)]{RLS01}
   Reid, I. N., Liebert, J., \& Schmidt, G. 2001, \apj, 550, L61
\bibitem[Reid, Kilkenny, \& Cruz(2002)]{RKC02}
   Reid, I. N., Kilkenny, D., \& Cruz, K. L. 2002, \aj, 123, 2822
\bibitem[Reyl\'e {\it et al.}(2002)]{RRSI02}
   Reyl\'e, C., Robin, A. C., Scholz, R.-D., \& Irwin, M. 2002, \aap,
   390, 491
\bibitem[Salim \& Gould(2002)]{SG02}
   Salim, S., \& Gould, A. 2002, \apj, 575, L83
\bibitem[Scholz {\it et al.}(2000)]{SIIJM00}
   Scholz, R.-D., Irwin, M., Ibata, R., Jahreiss, H., \& Malkov,
   O. Yu. 2000, \aap, 353, 958
\bibitem[Scholz {\it et al.}(2002a)]{SIILSS02a}
   Scholz, R.-D., Ibata, R., Irwin, M., Salvato, M., \& Schweitzer,
   A. 2002a, \mnras, 329, 109
\bibitem[Scholz {\it et al.}(2002b)]{SSAII02b}
   Scholz, R.-D., Szokoly, G. P., Andersen, M., Ibata, R., \& Irwin, M
   J. 2002b, \apj, 565, 539
\bibitem[Schweitzer {\it et al.}(1999)]{SSSIM99}
   Schweitzer, A., Scholz, R.-D., Stauffer, J., Irwin, M., \&
   McCaughrean, M. J. 1999, \aap, 350, L62
\bibitem[Wesemael(1993)]{WGLLFBG93}
   Wesemael, F., Greenstein, J. L., Liebert, James, Lamontagne, R.,
   Fontaine, G., Bergeron, P., \& Glaspey, J. W. 1993, \pasp, 105, 761
\end{thebibliography}
\end{document}